\documentclass[12pt,a4paper,notitlepage]{article}

\usepackage[a-2u]{pdfx}
\usepackage[utf8]{inputenc}
\usepackage{lmodern}
\usepackage{amsmath}        % extensions for typesetting of math
\usepackage{amsfonts}       % math fonts
\usepackage{amsthm}         % theorems, definitions, etc.
\usepackage{bbding}         % various symbols (squares, asterisks, scissors, ...)
\usepackage{bm}             % boldface symbols (\bm)
\usepackage{graphicx}       % embedding of pictures
\usepackage{fancyvrb}       % improved verbatim environment
\usepackage[nottoc]{tocbibind} % makes sure that bibliography and the lists
\usepackage{dcolumn}        % improved alignment of table columns
\usepackage{booktabs}       % improved horizontal lines in tables
\usepackage{paralist}       % improved enumerate and itemize
\usepackage{xcolor}         % typesetting in color
\usepackage{float}			% whips those tables back where they belong
\usepackage{titling}		% abstract on title page
\usepackage{cleveref}       %lepsi referovani
\usepackage{epsfig}         %eps obrazky
\usepackage[margin=2.5cm]{geometry}
\usepackage{multirow}
\usepackage{gensymb}        % adds degrees symbol
\usepackage{cite}
\usepackage{authblk}

\newcommand{\no}{\noindent}
\newcommand{\tcr}[1]{\textcolor{red}{#1}}

\title{Subthreshold parameters of $\pi\pi$ scattering revisited}
\author{Marián Kolesár\thanks{marian.kolesar@matfyz.cuni.cz} ~and Jaroslav Říha\thanks{jaroslav.riha@matfyz.cuni.cz}}
\affil{\footnotesize\it Institute of Particle and Nuclear Physics, Charles University, Prague, Czech Republic}
\makeatletter 				% thanks -> numbers
\let\@fnsymbol\@arabic
\makeatother

\begin{document}

\begin{titlingpage}
\maketitle

\begin{abstract}
Using a variety of experimental results and lattice QCD calculations of $\pi\pi$ scattering lengths, while employing dispersive representations of the amplitude based on Roy equations, we compute the subthreshold parameters of this process. We use Monte Carlo sampling to numerically model the probability distribution of the results based on all uncertainties in the inputs. We also investigate the dependence of the results on a theoretical correlation between the $\pi\pi$ scattering lengths $a^0_0$ and $a^2_0$, which was previously established in the framework of two-flavor $\chi$PT.
\end{abstract}
\end{titlingpage}

% \tableofcontents
\newpage

\section{Introduction}
\label{Sec:Introduction}
The $\pi\pi$ scattering amplitude can be constructed as a general solution of unitarity, analyticity and crossing symmetry up to and including $O(p^6)$ by employing the so-called reconstruction theorem \cite{Stern_1993, Stern_1995}

\begin{align}
\begin{split}
A_{\pi\pi}(s,t,u) = &\frac{\alpha_{\pi\pi}}{3F_{\pi}^{2}}M_{\pi}^{2}+\frac{\beta_{\pi\pi}}{3F_{\pi}^{2}}(3s-4M_{\pi}^{2})+\\
&+\frac{\lambda_{1}}{F_{\pi}^{4}}\left(s-2M_{\pi}^{2}\right)^{2}+\frac{\lambda_{2}}{F_{\pi}^{4}}\left[\left(t-2M_{\pi}\right)^{2}+\left(u-2M_{\pi}\right)^{2}\right]+\\
&+\frac{\lambda_{3}}{F_{\pi}^{6}}\left(s-2M_{\pi}^{2}\right)^{2}+\frac{\lambda_{4}}{F_{\pi}^{6}}\left[\left(t-2M_{\pi}\right)^{3}+\left(u-2M_{\pi}\right)^{3}\right]+\\
&+\overline{K}(s|t,u)+O\left(\frac{p^8}{\Lambda_{H}^8}\right),
\end{split} \label{alfabeta_pipi}
\end{align}

\noindent where $\overline{K}(s|t,u)$ contains unitarity corrections, the explicit form of which can be found in Appendix \ref{Kstu}. The coefficients $\alpha_{\pi\pi}$, $\beta_{\pi\pi}$, $\lambda_{1}\dots \lambda_{4}$ are called subthreshold parameters.

Such a representation of the amplitude is useful because good convergence of $\alpha_{\pi\pi}$ and $\beta_{\pi\pi}$ is expected within the framework of 3-flavor chiral perturbation theory ($\chi$PT) \cite{Gasser:1984gg}. As shown in \cite{Stern_2004}, the sum of next-to-next-leading order (NNLO) and all higher orders, written in a convenient form of the chiral expansion of these parameters, is proportional to
\begin{align}
    \delta_{\alpha_{\pi\pi}}^{NNLO} &\sim O(m_{ud}m_s) \\
    \delta_{\beta_{\pi\pi}}^{NNLO} &\sim O(m_{ud}m_s),
\end{align}

\no instead of $O(m_s^2)$, as is generally expected in $SU(3)$ $\chi$PT. Here, $m_{ud}$ denotes the average mass of the $u$ and $d$ quarks.

This feature was used by Descotes-Genon et al. \cite{Stern_2004} in an effort to restrict the values of three-flavor $\chi$PT low-energy constants (LECs) at the leading order. The analysis utilized the extraction of the subthreshold parameters by Descotes et al. (DFGS) \cite{DFGS}, which was based on values of $\pi\pi$ scattering lengths obtained primarily from experimental data by the BNL-E865 collaboration \cite{BNL-E865:2001wfj}:
\begin{equation}
    \alpha_{\pi\pi} = 1.381 \pm 0.242, \quad
    \beta_{\pi\pi} = 1.081 \pm 0.023 \quad \mathrm{(global\ fit)}. \label{Stern_alfa_pipi}
\end{equation}

\noindent In comparison, the values of both parameters at leading order (LO) are equal to one:
\begin{equation}
    \alpha_{\pi\pi}^{LO} = 1,\quad \beta_{\pi\pi}^{LO} = 1,
\end{equation}

\no which suggests a rather sizeable correction at next-to-leading order (NLO) for $\alpha_{\pi\pi}$. This leads to a suppressed pion mass at the leading order in the analysis by Descotes-Genon et al. \cite{Stern_2004}:

\begin{equation}
    Y(3) = \frac{(M_\pi^{LO})^2}{M_\pi^2}\ \sim\ 0,
\end{equation}

\no although with a large uncertainty rooted in the substantial error in the value of $\alpha_{\pi\pi}$ (\ref{Stern_alfa_pipi}). The square of the ratio of the leading-order pion mass and its physical value is connected to the leading-order LEC $B_0$

\begin{equation}
    Y(3) = \frac{2B_0 m_{ud}}{M_\pi^2}.
\end{equation}

On the other hand, Colangelo, Gasser and Leutwyler (CGL) \cite{CGL} obtained a significantly lower estimate of $\alpha_{\pi\pi}$ in the framework of two-flavor $\chi$PT.

\begin{equation}\label{eq:CGL_subthr}
    \alpha_{\pi\pi} = 1.08 \pm 0.07, \quad
    \beta_{\pi\pi} = 1.12 \pm 0.01.
\end{equation}

Furthermore, a combined analysis by Kolesár, Novotný \cite{Kolesar_2018} concluded that $\eta\to 3\pi$ data is in tension with the results of DFGS \cite{DFGS}. In particular, the source of this tension is the value of $\alpha_{\pi\pi}$ (\ref{Stern_alfa_pipi}) -- they considered this value suspicious in their investigation, as it prefers a very low value for the chiral condensate and thus $Y(3)$, which is not very consistent with current expectations from two-loop $\chi$PT fits \cite{Bijnens_2014}. For comparison, the result they obtained for $Y(3)$ purely from the $\eta\to 3\pi$ data was the following:

\begin{equation}
    Y(3)=1.44 \pm 0.32.
\end{equation}

\no The paper \cite{Kolesar_2018} concluded that a more precise determination of the value of the $\pi\pi$ subthreshold parameters, especially for $\alpha_{\pi\pi}$, is needed.

In fact, more recent information on the values of $\pi\pi$ scattering lengths is available. In this paper, we will take advantage of the experimental data published by the NA48/2 collaboration \cite{NA48}, which extracted the scattering lengths from the $K_{e4}$ decay and K$\to$3$\pi$ cusp measurements. We will also employ the ``Best values" fit-to-data quoted by Pelaez et al. \cite{Pelaez_2024}, which uses the parametrization originally developed by Garcia-Martin et al. (GKPRY) \cite{Garcia-Martin_2011}.

In addition, the scattering lengths have also been calculated by several lattice QCD groups, e.g., ETM \cite{ETM_a00,ETM_a20}, RBC/UKQCD \cite{RBC/UKQCD:2023pde}, Mai et al. \cite{Mai_2019}, Fu, Wang \cite{Fu_2024} and others \cite{NPLQCD_2012, PACS-CS_2014, Dawid:2025doq}. In the following, we will utilize the results by the ETM and RBC/UKQCD collaborations.

The goal of the paper is to extract the values of the subthreshold parameters from the recent data on the scattering lengths discussed above. This is achieved by employing dispersive representations of the amplitude based on Roy equations \cite{ACGL, DFGS}, following the analysis by the NA48/2 experiment \cite{NA48}. In the process, we also obtain new values of coefficients $\bar{b}_i$, used in an alternative representation of the $\pi\pi$ scattering amplitude \cite{Bijnens_1996, Bijnens:1997vq}. In our analysis, we numerically generate a large statistical ensemble in order to describe probability distributions, including various inputs and all uncertainties. Our best estimate combines both experimental and lattice QCD inputs.

In addition, we investigate the dependence of the results on a theoretical correlation between the $\pi\pi$ scattering lengths $a^0_0$ and $a^2_0$, employed by both the CGL analysis and the main result by the NA48/2 collaboration (model C), which can be established in the framework of two-flavor $\chi$PT and requires a phenomenological determination of the scalar radius of the pion \cite{CGL_korelace}. This correlation could be a potential source of a 1$\sigma$ discrepancy in the scattering lengths, as discussed by DFGS \cite{DFGS}, also visible in the values of $\alpha_{\pi\pi}$ mentioned above.

The paper is structured as follows. Section 2 introduces the $\pi\pi$ scattering amplitude. Section 3 outlines the procedure, which employs the solutions of the Roy equations in order to obtain the values of the subthreshold parameters and the coefficients $\bar{b}_i$ from the data on the scattering lengths. A detailed description of our inputs is then provided in Sec. 4, while Sec. 5 contains results and discussion. Finally, Sec. 6 summarizes the paper. Some of the explicit formulas and tables containing numerical inputs are postponed to the Appendixes.

\section{\boldmath $\pi\pi$ scattering amplitude} \label{Sec:pipi_scatt}

Elastic scattering of $\pi\pi$ in the isospin limit, with the EM interactions neglected, can be described by an amplitude $A_{\pi\pi}(s,t,u)$ (for more details, see ACGL \cite{ACGL})

\vspace{-0.4cm}
\begin{eqnarray}
    <\pi_{4}^d(p_{4})\pi_{3}^c(p_{3})| \pi_{1}^a(p_{1})\pi_{2}^b(p_{2})>\ =\ 
	\delta_{fi}+(2\pi)^{4}i\delta^{(4)}(P_{f}-P_{i})\cdot \hspace{2.5cm} \nonumber\\ 
	\cdot [\delta^{ab}\delta^{cd}A_{\pi\pi}(s,t,u) + \delta^{ac}\delta^{bd}A_{\pi\pi}(t,u,s) + 
	\delta^{ad}\delta^{bc}A_{\pi\pi}(u,s,t)]. 
\end{eqnarray}

\noindent with the $s$-channel isospin components defined as
\begin{eqnarray}
T^{0}(s,t)&=&3A_{\pi\pi}(s,t,u)+A_{\pi\pi}(t,u,s)+A_{\pi\pi}(u,s,t)\\
T^{1}(s,t)&=&A_{\pi\pi}(t,u,s)-A_{\pi\pi}(u,s,t)\\
T^{2}(s,t)&=&A_{\pi\pi}(t,u,s)+A_{\pi\pi}(u,s,t).
\end{eqnarray}

\noindent The partial-wave decomposition can be introduced in the following way
\begin{eqnarray}
T^{I}(s,t)&=&32\pi\sum_{l}{(2l+1)P_{l}\left(1+\frac{2t}{s-4M_{\pi}^{2}}\right)t_{l}^{I}(s)}\\
t_{l}^{I}(s)&=&\frac{1}{2i\sigma(s)}\left(\eta_{l}^{I}(s)e^{2i\delta_{l}^{I}(s)}-1\right)\\
\mathrm{Re} t_{l}^{I}(s)&=&q^{2l}\left(a_{l}^{I}+b_{l}^{I}q^{2}+c_{l}^{I}q^{4}+...\right)\\
\sigma(s)&=&\sqrt{1-\frac{4M_{\pi}^{2}}{s}},
\end{eqnarray}

\noindent where $q=\frac{1}{2}\sqrt{s-4M_{\pi}^2}$.

According to the Roy equations \cite{Roy_1971}, which are a consequence of analyticity, unitarity, crossing symmetry, and asymptotic bounds, it is possible to represent the low-energy $\pi\pi$ amplitude in terms of only two subtraction constants in a suitable kinematical range ($\sqrt{s} < \sqrt{s_0} = 800$ MeV). These can be identified as the scattering lengths $a_{0}^{0}$ and $a_{0}^{2}$

\begin{align}
\label{phenomenologicalrepre}
\begin{split}
A_{\pi\pi}(s,t,u) = &16\pi a_{0}^{2}+\frac{4\pi}{3M_{\pi}^{2}}\left(2a_{0}^{0}-5a_{0}^{2}\right)s+P(s,t,u)+32\pi[\frac{1}{3}\overline{W}^{0}(s)+\\
&\frac{3}{2}(s-u)\overline{W}^{1}(t)+\frac{3}{2}(s-t)\overline{W}^{1}(u)+\frac{1}{2}\overline{W}^{2}(u)+\frac{1}{2}\overline{W}^{2}(u)-\\
&\frac{1}{3}\overline{W}^{2}(s)]+O(p^{8}).
\end{split}
\end{align}

\no The dispersion integrals are defined as

\begin{eqnarray}
\overline{W}^{0}(s)&=&\frac{s^{4}}{\pi}\int_{4M_{\pi}^{2}}^{\infty}{\frac{\mathrm{Im} t_{0}^{0}(s')}{s'^{4}(s'-s)}ds'}\\
\overline{W}^{1}(s)&=&\frac{s^{3}}{\pi}\int_{4M_{\pi}^{2}}^{\infty}{\frac{\mathrm{Im} t_{1}^{1}(s')}{s'^{3}(s'-M_{\pi}^{2})(s'-s)}ds'}\\
\overline{W}^{2}(s)&=&\frac{s^{4}}{\pi}\int_{4M_{\pi}^{2}}^{\infty}{\frac{\mathrm{Im} t_{2}^{0}(s')}{s'^{4}(s'-s)}ds'}
\end{eqnarray}

\no and the crossing symmetry polynomial $P(s,t,u)$ is
\begin{eqnarray}
P(s,t,u)=p_{1}+p_{2}s+p_{3}s^{2}+p_{4}(t-u)^{2}+p_{5}s^{3}+p_{6}s(t-u)^{2}.
\end{eqnarray}

\no The coefficients $p_{i}$ can be expressed as
\begin{eqnarray}
p_{1}&=&-128\pi M_{\pi}^{4}\left(\overline{I}_{0}^{1}+\overline{I}_{0}^{2}+2M_{\pi}^{2}\overline{I}_{1}^{1}+2M_{\pi}^{2}\overline{I}_{1}^{2}+8M_{\pi}^{4}\overline{I}_{2}^{2}\right)\\ \nonumber
p_{2}&=&\frac{-64\pi M_{\pi}^{2}}{3}\left(2\overline{I}_{0}^{0}-6\overline{I}_{0}^{1}-\overline{I}_{0}^{2}-15M_{/pi}^{2}\overline{I}_{1}^{1}-3M_{\pi}^{2}\overline{I}_{1}^{2}-36M_{\pi}^{4}\overline{I}_{2}^{2}+6M_{\pi}^{2}H\right)\\ \nonumber
p_{3}&=&\frac{8\pi}{3}\left(4\overline{I}_{0}^{0}-9\overline{I}_{0}^{1}-\overline{I}_{0}^{2}-16M_{\pi}^{2}\overline{I}_{1}^{0}-42M_{\pi}^{2}\overline{I}_{1}^{1}+22M_{\pi}^{2}\overline{I}_{1}^{2}-72M_{\pi}^{4}\overline{I}_{2}^{2}+24M_{\pi}^{2}H\right)\\ \nonumber
p_{4}&=&8\pi\left(4\overline{I}_{0}^{1}+\overline{I}_{0}^{2}+2M_{\pi}^{2}\overline{I}_{1}^{1}+2M_{\pi}^{2}\overline{I}_{1}^{2}-24M_{\pi}^{4}\overline{I}_{2}^{2}\right)\\ \nonumber
p_{5}&=&\frac{4\pi}{3}\left(8\overline{I}_{1}^{0}+9\overline{I}_{1}^{1}-11\overline{I}_{1}^{2}-32M_{\pi}^{2}\overline{I}_{2}^{0}+44M_{\pi}^{2}\overline{I}_{2}^{2}-6H\right)\\ \nonumber
p_{6}&=&4\pi\left(\overline{I}_{1}^{1}-3\overline{I}_{1}^{2}+12M_{\pi}^{2}\overline{I}_{2}^{2}+2H\right), \nonumber\label{eq:p_i}
\end{eqnarray}

\noindent with the moments $\overline{I}_{n}^{I}$ and $H$ being
\begin{eqnarray}
\label{eq:integrals} \overline{I}_{n}^{I}&=&\sum_{l=0}^{\infty}\frac{2l+1}{\pi}\int_{4M_{\pi}^{2}}^{\infty}{\frac{\mathrm{Im} t_{l}^{I}(s)}{s^{n+2}(s-4M_{\pi}^{2})}}ds\\ 
H&=&\sum_{l=2}^{\infty}{\frac{(2l+1)l(l+1)}{\pi}\int_{4M_{\pi}^{2}}^{\infty}{\frac{2\mathrm{Im} t_{l}^{0}(s)+4\mathrm{Im} t_{l}^{2}(s)}{9s^{3}(s-4M_{\pi}^{2})}}}ds.
\end{eqnarray}

Following ACGL \cite{ACGL} and CGL \cite{CGL}, we decompose the amplitude

\begin{align}
    A_{\pi\pi}(s,t,u)=A_{\pi\pi}^{SP}(s,t,u)+A_{\pi\pi}^{d}(s,t,u),\label{amplitude_split}
\end{align}

\noindent where $A_{\pi\pi}^{SP}(s,t,u)$ describes the contributions generated by the imaginary parts of the S and P waves below $\sqrt{s_2}=2$ GeV and $A_{\pi\pi}^{d}(s,t,u)$ contains higher energies as well as higher partial waves.

Corresponding to (\ref{amplitude_split}), the moments are split 

\begin{align}
    \label{eq:I_moments}
    \nonumber \overline{I}^0_n &= J^0_n + I^0_n, \qquad \overline{I}^1_n = 3J^1_n + I^1_n, \qquad \overline{I}^2_n = J^2_n + I^2_n \\
    \nonumber J^0_n&=\frac{1}{\pi}\int_{4M_{\pi}^{2}}^{s_2}{\frac{\mathrm{Im} t_{0}^{0}(s)}{s^{n+2}(s-4M_{\pi}^{2})}}ds\\
    J^1_n&=\frac{1}{\pi}\int_{4M_{\pi}^{2}}^{s_2}{\frac{\mathrm{Im} t_{1}^{1}(s)}{s^{n+2}(s-4M_{\pi}^{2})}}ds\\
    \nonumber J^2_n&=\frac{1}{\pi}\int_{4M_{\pi}^{2}}^{s_2}{\frac{\mathrm{Im} t_{0}^{2}(s)}{s^{n+2}(s-4M_{\pi}^{2})}}ds\\
    \nonumber I_{n}^{I}&=\sum_{l=2}^{\infty}\frac{2l+1}{\pi}\int_{4M_{\pi}^{2}}^{s_2}{\frac{\mathrm{Im} t_{l}^{I}(s)}{s^{n+2}(s-4M_{\pi}^{2})}}ds+\sum_{l=0}^{\infty}\frac{2l+1}{\pi}\int_{s_2}^{\infty}{\frac{\mathrm{Im} t_{l}^{I}(s)}{s^{n+2}(s-4M_{\pi}^{2})}}ds, 
\end{align}

\noindent where the partial-wave amplitudes may be expressed in terms of phase shifts $\delta_{l}^{I}(s)$

\begin{eqnarray}
    t_{l}^{I}(s)=\frac{1}{\sigma(s)}e^{i\delta_{l}^{I}(s)}\sin(\delta_{l}^{I}(s)). 
    \label{royamplitude}
\end{eqnarray}

We will compute $J^I_n$ explicitly, using solutions of Roy equations by ACGL \cite{ACGL} and DFGS \cite{DFGS}, while we will use the numerical estimates for the moments of the background amplitude $I^I_n$ and $H$ from ACGL \cite{ACGL}.

These numerical solutions to the Roy equations (\cite{ACGL} and \cite{DFGS}) employ the so-called Schenk parametrization

\begin{eqnarray}
\tan(\delta_{l}^{I})=\sqrt{1-\frac{4M_{\pi}^{2}}{s}}q^{2l}\left(A_{l}^{I}+B_{l}^{I}q^{2}+C_{l}^{I}q^{4}+D_{l}^{I}q^{6}\right)\frac{4M_{\pi}^{2}-s_{l}^{I}}{s-s_{l}^{I}}, \label{tandelta}
\end{eqnarray}

\noindent where $A_{l}^{I},B_{l}^{I},C_{l}^{I},D_{l}^{I},s_{l}^{I}$ are the Schenk parameters, which are approximated by a third-degree polynomial in scattering lengths. Using $A_{0}^{0}$ as an example
\begin{equation} \label{eq_schenkpar}
A_{0}^{0}=z_{1}+z_{2}v+z_{4}u^{2}+z_{5}v^{2}+z_{6}uv+z_{7}u^{3}+z_{8}u^{2}v+z_{9}uv^{2}+z_{10}v^{3},
\end{equation}
where \emph{u} and \emph{v} are
\begin{align}
u&=\frac{a_{0}^{0}}{p_{0}}-1 \hspace{1cm} v=\frac{a_{0}^{2}}{p_{2}}-1,\\
p_{0}&=0.225 \hspace{1.1cm} p_{2}=-0.03706.
\end{align}

The extended solution by DFGS \cite{DFGS} also considers the Schenk parameters to be explicitly dependent on the uncertainty in the phase shift at the matching point $\sqrt{s_0}=800$ MeV, such that 
\begin{equation} \label{eq_schenkpar_z}
    z_{j}=a_{j}+\delta \theta_{0} b_{j}+\delta\theta_{1} c_{j}
\end{equation}
with 
\begin{align}
    \delta\theta_{0}&=\theta_{0}-82.3\degree \hspace{1.2cm} \theta_0=82.3\degree \pm 3.4\degree,\\
    \delta\theta_{1}&=\theta_{1}-108.9\degree \hspace{1cm}\theta_1=110.4\degree \pm 0.7\degree,
\end{align}

\noindent where for $\theta_1$ we use the improved determination by \cite{Colangelo:2018mtw}. The values for $\theta_0$ and $\theta_1$ are compatible within uncertainties with fits-to-data of the phase shifts in \cite{Garcia-Martin_2011,Pelaez_2019,Pelaez_2024}.

In this approach, the parameters $s_{0}^{0}$, $s_{1}^{1}$ and $s_{0}^{2}$ are fixed by the boundary conditions
\begin{align}
    \delta_{0}^{0}\left(s_{0}\right)&\equiv \theta_{0}\\
    \delta_{1}^{1}\left(s_{0}\right)&\equiv \theta_{1}\\
    \delta_{0}^{2}\left(s_{0}\right)&\equiv \theta_{2},
\end{align}

\noindent instead of being given by fixed values, as in ACGL \cite{ACGL}. Here, $\theta_{2}\left(a_{0}^{0},a_{0}^{2},\theta_{0},\theta_{1}\right)$ is parametrized analogously to the Schenk parameters by Eqs. (\ref{eq_schenkpar}) and (\ref{eq_schenkpar_z}). The explicit solutions can be found in Appendix \ref{schenkpar}.

This framework uniquely fixes the three relevant phase shifts at
low energies in terms of the two scattering lengths $a^0_0$ and $a^2_0$, and experimental data above $s_0$. This allows for a model-independent determination of the scattering lengths from the phase shifts at energies $s < s_0$, within the so-called universal band allowed by the high-energy constraints \cite{ACGL}. Conversely, information about the two $S$-wave scattering lengths makes it possible to determine any other low-energy $\pi\pi$ scattering observable, for example, the subthreshold parameters, the subject of interest in this article.

The phenomenological representation introduced above was used, for the purpose of extracting the scattering lengths, by the NA48/2 collaboration \cite{NA48}, which is our main experimental input. As will be discussed later, our calculations in this article will primarily use the extended solution by DFGS \cite{DFGS}.

It should be noted that a large amount of work regarding the $\pi\pi$ scattering amplitude is available in the literature \cite{Kaminski_2002, Caprini_2011, Moussallam_2011, Garcia-Martin_2011, Albaladejo_2018, Pelaez_2019, Niehus_2020, Cao_2023, Pelaez_2024, Guerrieri_2024}. As an example, Caprini et al. \cite{Caprini_2011} adopted Regge parametrization in order to describe the $\pi\pi$ amplitude at higher energies. Cao et al. \cite{Cao_2023} extended the Roy equations using unphysical pion masses. Pelaez et al. \cite{Pelaez_2019} provide a global parametrization valid up to 2 GeV, while Pelaez et al. \cite{Pelaez_2024} extend the parametrization to higher partial waves. Recently, Guerrieri et al.\cite{Guerrieri_2024} have claimed to obtain the scattering lengths with high precision using a Bootstrap fit of the $\pi\pi$ scattering amplitude, but the work has not been fully published yet.

One can ask if employing a more advanced representation, such as the one from \cite{Pelaez_2024}, might not be advantageous in our case in addition to directly using the results from the NA48/2 collaboration \cite{NA48}. Such an application would be very straightforward if it were possible to interpret the subthreshold representation as an expansion of the amplitude at a specific kinematic point under threshold. Unfortunately, this is not really the case, as can be seen from the definition (\ref{alfabeta_pipi}). It would be necessary to develop a matching of (\ref{alfabeta_pipi}) to the considered extended representation, which is beyond the scope of our current article. What we can do, however, is to utilize the available matching of the phenomenological and subthreshold representations, discussed in the following Section, and use imaginary parts of the partial waves and the extracted scattering lengths as inputs. We will take the "Best values" fit quoted in \cite{Pelaez_2024}, which uses the parametrization originally developed by Garcia-Martin et al. (GKPRY) \cite{Garcia-Martin_2011} (CFD -- constrained fit-to-data).

\section{Extraction of subthreshold parameters from scattering lengths} \label{sec:subthr_extraction}

As discussed in the Introduction, the amplitude can also be constructed as a general solution of unitarity, analyticity, and crossing symmetry \cite{Stern_1995, Kampf_2019}. Hence, the subthreshold parameters $\alpha_{\pi\pi}$, $\beta_{\pi\pi}$, $\lambda_{1}\dots \lambda_{4}$ can be introduced\cite{Stern_1995}, which we would like to extract from data for the scattering lengths. Therefore, a way to relate the subthreshold representation (\ref{alfabeta_pipi}) with the phenomenological one (\ref{phenomenologicalrepre}) is needed.

For this reason, we will briefly mention the chiral representation \cite{CGL}, utilizing coefficients $c_{i}$ (defined by the relations in Appendix \ref{App_repre_rel}), and $\overline{b}_{i}$ representation, introduced by Bijnens et al. \cite{Bijnens_1996} and formulated as follows:

\begin{align}
\begin{split}
A_{\pi\pi}(s,t,u) = &\frac{M_{\pi}^{2}}{F_{\pi}^{2}}\left(s-1\right)+\frac{M_{\pi}^{4}}{F_{\pi}^{4}}\left(\overline{b}_{1}+\overline{b}_{2}s+\overline{b}_{3}s^{2}+\overline{b}_{4}\left(t-u\right)^{2}\right)+\\
&+\frac{M_{\pi}^{4}}{F_{\pi}^{4}}\left(F^{(1)}(s)+G^{(1)}(s,t)+G^{(1)}(s,u)\right)+\\
&+\frac{M_{\pi}^{6}}{F_{\pi}^{6}}\left(\overline{b}_{5}s^{3}+\overline{b}_{6}s\left(t-u\right)^{2}\right)+\\
&+\frac{M_{\pi}^{6}}{F_{\pi}^{6}}\left(F^{(2)}(s)+G^{(2)}(s,t)+G^{(2)}(s,u)\right)+O\left(\left(\frac{M_{\pi}^{2}}{F_{\pi}^{2}}\right)^{4}\right),
\end{split}
\end{align}

\noindent where the nonpolynomial part can be found in Appendix \ref{App_b_representation}. This representation was used in the context of calculating two-loop corrections to $\pi\pi$ scattering within two-flavor $\chi$PT \cite{Bijnens:1997vq}. 

The subthreshold parameters can be related to the $\overline{b}_{i}$--representation as \cite{CGL}
\begin{eqnarray} \label{eq:alfafrombi}
\nonumber \alpha_{\pi\pi}&=&1+\xi(3\overline{b}_{1}+4\overline{b}_{2}+4\overline{b}_{3}-4\overline{b}_{4})-\frac{11}{36}\pi^{2}\xi^{2}-\frac{152}{9}\xi^{2}\\
\nonumber \beta_{\pi\pi}&=&1+\xi(\overline{b}_{2}+4\overline{b}_{3}-4\overline{b}_{4})+4\xi^{2}(3\overline{b}_{5}-\overline{b}_{6})-\frac{13}{72}\pi^{2}\xi^{2}+\frac{152}{9}\xi^{2} \label{betafrombi}\\
\nonumber N\lambda_{1}&=& \overline{b}_{3}-\overline{b}_{4}+2\xi\left(3\overline{b}_{5}-\overline{b}_{6}\right)+\frac{\pi^{2}}{48}\xi+\frac{38}{3}\xi\\
N\lambda_{2}&=&2\overline{b}_{4}-\frac{1}{3}\xi\\
\nonumber N^{2}\lambda_{3}&=&\overline{b}_{5}-\frac{1}{3}\overline{b}_{6}+\frac{82}{27}\\
\nonumber N^{2}\lambda_{4}&=&-\frac{4}{3}\overline{b}_{6}-\frac{14}{27},
\end{eqnarray}

\noindent where $\xi=\left(\frac{M_{\pi}}{4\pi F_{\pi}}\right)^{2}$ and $N=16\pi^{2}$.

To calculate $\alpha_{\pi\pi}$, $\beta_{\pi\pi}$ and $\lambda_i$ from known values of scattering lengths, we first need to compute the moments $\overline{I}_{l}^{I}$ in Eq.(\ref{eq:I_moments}). Then we obtain $p_{i}$'s from the phenomenological representation using Eqs. (\ref{eq:p_i}) and transform them into $c_{i}$'s from the chiral representation by employing (\ref{ci_pi}) (see Appendix \ref{App_repre_rel}). In the next step, we use the inverse of the formula (\ref{ci_bi}) to transform $c_{i}$ into $\overline{b}_{i}$'s and subsequently into the subthreshold parameters using (\ref{eq:alfafrombi}). 

The procedure, adopted from DFGS \cite{DFGS}, can be outlined as follows:
\begin{table}[H]
\centering
$a_{0}^{0}$ and $a_{0}^{2}$ from exp./lattice data + solutions to the Roy equations\\
$\downarrow$\\
phase shifts\\
$\downarrow$\\
imaginary parts of partial wave amplitudes\\
$\downarrow$\\
moments of background amplitude$\left(\overline{I}_{i}^{j} \ \mathrm{'s}\right)$\\
$\downarrow$\\
phenomenological representation ($p_{i}$'s)\\
$\downarrow$\\
chiral representation ($c_{i}$'s)\\
$\downarrow$\\
$\overline{b}_{i}$'s\\
$\downarrow$\\
$\alpha_{\pi\pi}$, $\beta_{\pi\pi}$ and $\lambda_i$
\end{table}

We implement the procedure numerically, with all uncertainties being modeled by Monte Carlo sampling. We generate $10^5$ entries for the inputs $a_0^0$, $a_0^2$, $\theta_0$, $\theta_1$, $F_\pi$, and the background moments $I_n^I$ and $H$. The tiny experimental error in the value of the pion mass is neglected. The procedure is followed for all entries from each set and we thus obtain a statistical ensemble of values for the parameters $\overline{b}_{i}$, $\alpha_{\pi\pi}$, $\beta_{\pi\pi}$ and $\lambda_i$, capturing the final uncertainty generated from the inputs.

\section{\boldmath Data on $\pi\pi$ scattering lengths}\label{subsec:data}

As mentioned in the Introduction, our inputs come from experimental data and lattice QCD calculations. For the experimental part, we will use the main result by the NA48/2 collaboration (model C) \cite{NA48}, which obtained the scattering lengths by combining measurements of the $K_{e4}$ decay and the cusp in $K\to 3\pi$. Model C relies on a correlation between $a_0^0$ and $a_0^2$, which is based on the relation between the scalar pion radius and the scattering lengths \cite{CGL, CGL_korelace}:
\begin{equation} \label{eq:a20}a_0^2=-0.0444+0.236(a_0^0-0.22)-0.61(a_0^0-0.22)^2-9.9(a_0^0-0.22)^3 \pm 0.0008.
\end{equation}

\no This constraint can be established in the framework of two-flavor $\chi$PT, which requires a phenomenological determination of the scalar radius of the pion \cite{CGL}
\begin{equation}
    \langle r^2 \rangle_s = 0.61 \pm 0.04\ \mathrm{fm}^2.
\end{equation}

In addition, we will utilize the imaginary
parts of the partial waves and the scattering lengths extracted by GKPRY \cite{Garcia-Martin_2011}, as discussed in Sec. \ref{Sec:pipi_scatt}. This work developed a representation of the $\pi\pi$ scattering amplitude up to 1.4 GeV. They fitted it to a range of experimental data, including the phase shifts at low energies measured by the NA48/2 collaboration in the $K_{e4}$ decay mentioned above. As our procedure explicitly depends on the scattering lengths (see \ref{ci_pi}), we have chosen the ``Best values" fit, because it has the smallest uncertainties for the scattering lengths, as evaluated by \cite{Pelaez_2024}.

Regarding the lattice QCD inputs, we will use the most recent calculation by the RBC/UKQCD group \cite{RBC/UKQCD:2023pde} and results by the ETM collaboration \cite{ETM_a00,ETM_a20}. The ETM calculation of $a_{0}^{2}$ \cite{ETM_a20} is highly rated by the Flavour Lattice Averaging Group (FLAG) \cite{FLAG21} and has a low associated error bar. However, it has been argued \cite{Draper:2021wga} that ETM's twisted-mass lattice QCD calculations may have a large systematic error in the case of $a_{0}^{0}$, unaccounted for in the paper \cite{ETM_a00}.

We will also try to reproduce the results cited by CGL \cite{CGL}, calculated in the framework of 2-flavor $\chi$PT, and DFGS \cite{DFGS}, based primarily on phase shift data from the BNL-E865 collaboration. Table \ref{table:scatter_len} contains all the published data on the scattering lengths discussed above.

\begin{table}[H]
    \centering
    \begin{tabular}{|c||c|c|}
        \hline \rule[0cm]{1cm}{0cm}
        &$a_0^0$&$a_0^2$\\
        \hline\hline
        NA48/2\cite{NA48} (model C) & $0.2196\pm 0.0034$ & Eq. (\ref{eq:a20})\\
        \hline
        GKPRY \cite{Garcia-Martin_2011} (Best values) & $0.220\pm 0.008$ &  $-0.042\pm 0.004$ \\
        \hline
        RBC/UKQCD \cite{RBC/UKQCD:2023pde} & $0.218\pm 0.034$ & $-0.0477\pm 0.0048$ \\
        \hline
        ETM collaboration \cite{ETM_a00,ETM_a20} & $0.198\pm0.011$ & $-0.0442(2)(\substack{+4 \\ -0})$\\
        \hline\hline
        CGL \cite{CGL} & $0.220\pm 0.005$ & Eq. (\ref{eq:a20})\\
        \hline
        DFGS \cite{DFGS} (Extended fit) & $0.228\pm 0.013$ &$-0.0380\pm 0.0044$ \\
        \hline
    \end{tabular}
    \caption{Inputs for scattering lengths used in this article in comparison with  CGL and DFGS. The values for DFGS are for their extended fit, which includes input values for the phase shifts $\theta_0$, $\theta_1$ at the matching point and a correlation matrix. Explicit values can be found in \cite{DFGS}.}
    \label{table:scatter_len}
\end{table}

As can be seen, the systematic uncertainty in the case of the ETM calculation of $a_{0}^{2}$ \cite{ETM_a20} is highly asymmetric. In the following, we modeled the combined distribution including both the statistical and systematic uncertainties as a skewed normal distribution with location $\mu=-0.04420$, scale $\sigma=0.00045$ and shape $\alpha=2$. This is numerically very close to a combination of a normal distribution with mean $\mu=-0.0442$ and standard deviation $\sigma=0.0002$, describing the statistical error, and a half-normal distribution with $\sigma=0.0004$, corresponding to the reported systematic uncertainty.

The articles \cite{DFGS, CGL} discuss the importance of the assumption (\ref{eq:a20}). The CGL paper \cite{CGL} explicitly relies on it, while DFGS \cite{DFGS} employ three different fits. The one listed in Table \ref{table:scatter_len} is the so-called ``extended" fit, which is based on the extended solution discussed in Section \ref{Sec:pipi_scatt} and does not use the assumption (\ref{eq:a20}). The authors considered their 'global' fit as the main result, but this does not take into account the uncertainty in the phase shifts at the matching point $s_0$. Furthermore, for comparison purposes, the authors also developed a ``scalar" fit, which employs the CGL correlation (\ref{eq:a20}).

One of the objectives of our paper is to test the sensitivity of the results on the assumption (\ref{eq:a20}). It should be stressed that the numerical solutions to the Roy equations that we employ in this paper \cite{ACGL, DFGS} are not dependent on this correlation, which is based on the determination of the scalar radius of the pion. Therefore, we utilized the independent determination of $a_0^2$ by ETM \cite{ETM_a20} in addition to the experimental measurement by the NA48/2 collaboration. While it would be straightforward to use the published $S0$-wave phase shifts from the $K_{e4}$ decay by NA48/2, our aim is to take advantage of the enhanced precision of the combined $K_{e4}$ and $K\to3\pi$ cusp results, which are only available for the scattering lengths. That is the reason why we have to infer values of phase shifts at low energies as an intermediary step.

Our procedure is the following: we used the measured scattering lengths by NA48/2 (model C) to reconstruct the phase shift $\delta_0^0$ at energies $s=\{4.2, 5, 6, 7, 8\}M^2_{\pi}$ ($10^4$ random generated entries for each energy) using the ACGL Schenk parametrization (Appendix \ref{schenkpar}). This low-energy region of center-of-mass system energies ($2M_{\pi}$ -- 400 MeV) is below the matching point $s_0$ and corresponds to the range probed in the $K_{e4}$ measurement by NA48/2. Then Eq. (\ref{tandelta}) is solved for each of the entries using the ETM distribution for $a_0^2$ as input, thus obtaining an ensemble of $a_0^0$ values, now independent of the correlation (\ref{eq:a20}). The obtained statistical ensemble has the following distribution  

\vspace{-0.25cm}
\begin{align}\label{eq:etm_refit}
        a_0^0 = 0.2203 (28), \quad a_0^2 = -0.0439 (3), \quad 
        \rho = 0.13 \qquad \mathrm{(NA48/2+ETM\ refit)}
\end{align}

\no $a_0^2$ is shown as symmetric here for the sake of simplicity, but the ensemble has the asymmetric distribution, as described above.

Due to the aforementioned associated systematic error in the ETM calculation of $a_0^0$, we also use a combination of the ETM input for $a^2_0$ and the CGL correlation (\ref{eq:etm_CGL_corr}) as an additional option:

\vspace{-0.25cm}
\begin{align}\label{eq:etm_CGL_corr}
    a_0^0& = 0.2223 (37), \quad a_0^2 = -0.0439 (3), \quad
    \rho = 0.37 \qquad \mathrm{(ETM + CGL\ corr.).}
\end{align}

In this paper, we use input from PDG \cite{pdg} for the pion decay constant:
\begin{align}
    F_{\pi}&=\frac{1}{\sqrt{2}}(130.56\pm 0.14)\ \mathrm{MeV}.
\end{align}

\no As for the moment $H$, we use the value from ACGL \cite{ACGL}:
\begin{align}
H=0.32\pm 0.02 \ \mathrm{GeV}^{-6}. \nonumber
\end{align}

\section{Results} \label{resultssection}

\begin{figure}[H]
    \centering
    \includegraphics[scale=0.54]{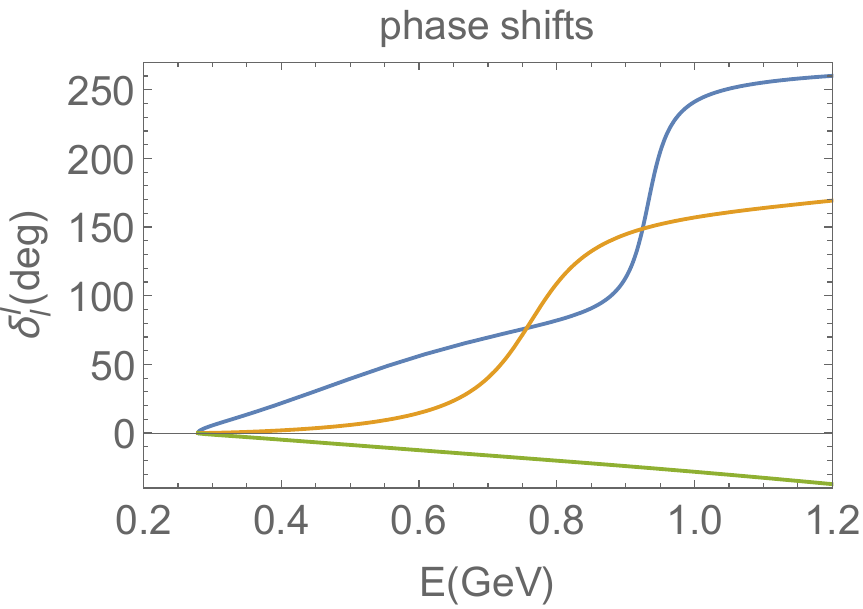} 
    \includegraphics[scale=0.54]{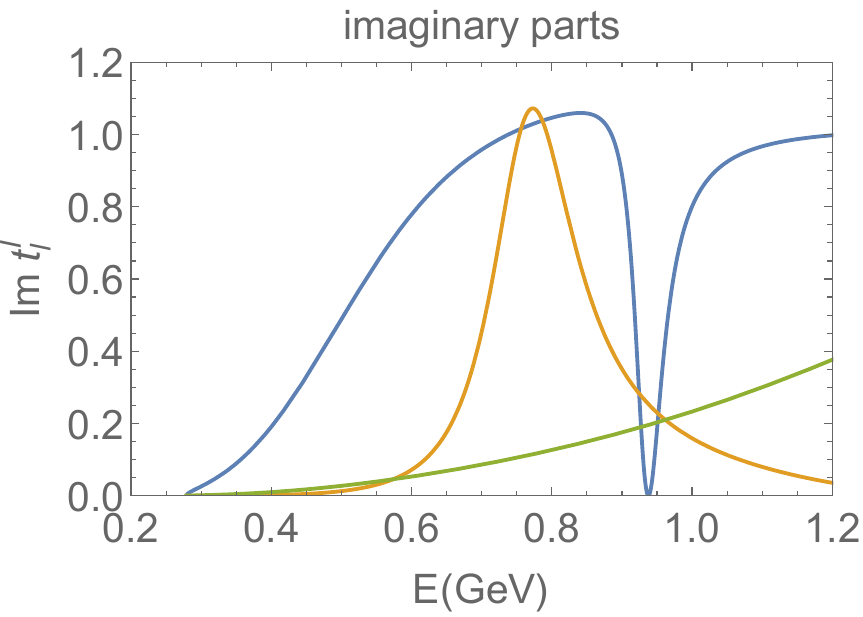} 
    \caption{Illustration of reconstructed phase shifts and imaginary parts of the partial-wave amplitudes at $a_{0}^{0}=0.2196$, $a_{0}^{2}=-0.0444$, using the ACGL solution \cite{ACGL}. Blue -- $t_0^0$, orange -- $t_1^1$, green -- $t_0^2$.}
    \label{phaseshiftfig}
\end{figure}

Following the procedure outlined in Sec. \ref{sec:subthr_extraction}, we prepared a numerical ensemble of $10^5$ entries for the inputs $a_0^0$, $a_0^2$, $\theta_0$, $\theta_1$, $F_\pi$ and the background moments $I_n^I$ and $H$. We first calculated the phase shifts $\delta_{0}^{0}$, $\delta_{1}^{1}$ and $\delta_{0}^{2}$ using Eq. (\ref{tandelta}). The dependence of the phase shifts and the imaginary parts of the corresponding partial wave amplitude on energy is illustrated in Fig. \ref{phaseshiftfig}.

The procedure is slightly different in the case of employing the inputs by GKPRY \cite{Garcia-Martin_2011}. In this case, we directly calculate the imaginary parts of the partial wave amplitudes from their representation, as given in \cite{Garcia-Martin_2011}.

\begin{figure}[H]
   \centering
    \includegraphics[scale=0.85]{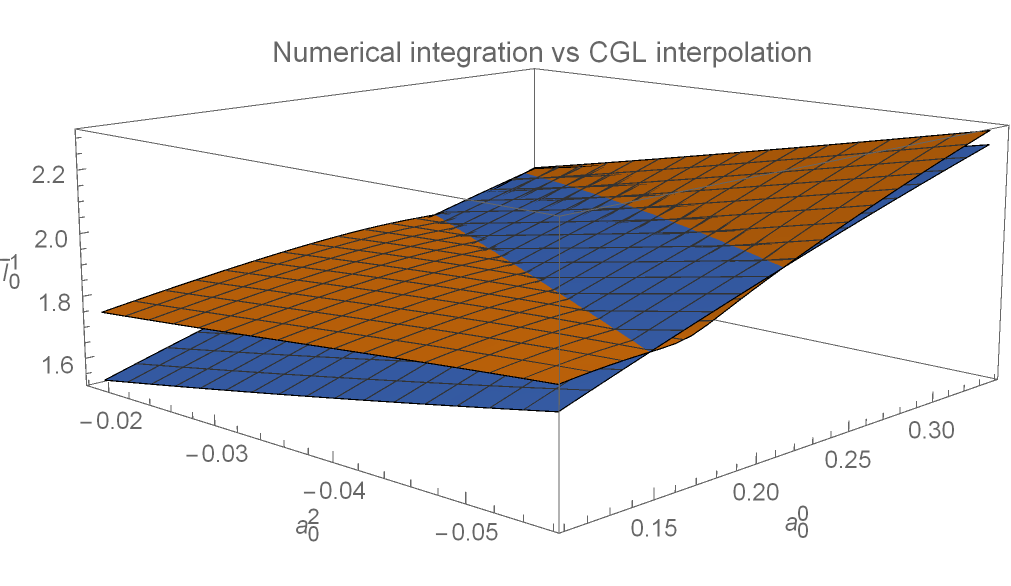}
    \caption{Moment $\overline{I}_0^1$: comparison of our \textcolor{orange}{numerical integration} and \textcolor{blue}{CGL's} \cite{CGL} quadratic interpolation as a function of the scattering lengths.}
    \label{fig:I10} 
\end{figure}

Next, we perform the integrations in (\ref{eq:I_moments}) numerically and obtain the moments $\overline{I}_k^j$. Figure \ref{fig:I10} showcases the dependence of one of the moments on the scattering lengths. One can observe a difference between the quadratic interpolation used by CGL \cite{CGL} and the explicit numerical integration that we used. However, the difference in the physical region is minimal.

Having calculated $\overline{I}^j_k$, we compute $p_i$'s from Eqs. (\ref{eq:p_i}) and match this representation with the chiral representation (for more details see Appendix \ref{App_repre_rel}). Finally, we get the statistical ensemble of values for the $\overline{b}_{i}$ coefficients  and the subthreshold parameters using Eqs. (\ref{eq:alfafrombi}).

For comparison purposes, we tried to reproduce the results from CGL \cite{ACGL} and DFGS \cite{DFGS}, using our methodology with their published inputs. In order to take into account the uncertainty in the phase shifts at the matching point $s_0$, we used the extended solution as discussed in Sec. \ref{Sec:pipi_scatt}. Although we were unable to get an exact match, our results were in general agreement with a few exceptions -- we obtained somewhat larger error bars for $\bar{b}_1$ in the case of CGL and for $\bar{b}_4$ and $\lambda_2$ in the case of DFGS.

Our results for the inputs in Table \ref{table:scatter_len} can be found in Table \ref{table:bi}. As can be seen, the obtained values are compatible with the two-flavor $\chi$PT predictions of CGL and, in the case of NA48/2 and ETM, the uncertainties are competitive. The error bars are significantly reduced for some of the parameters in comparison with the DFGS extended fit. In particular, this is the case for $\bar{b}_1$ and $\alpha_{\pi\pi}$, for which we can also exclude the large mean values suggested by DFGS.

Although compatible, the results based on the GKPRY CFD inputs exhibit larger uncertainties for the majority of the parameters compared to the NA48/2 case. This can be explained by the larger error bars in their determination of the scattering lengths, as can be observed in Table \ref{table:scatter_len}. The NA48/2 collaboration measured the threshold parameters with unique precision using the experimental combination of $K_{e4}$ decay and $K \to 3 \pi$ cusp measurements, and the theoretical constraint (\ref{eq:a20}). As follows from their analysis \cite{NA48}, results based solely on the phase shifts extracted from the $K_{e4}$ decay do not seem to achieve such accuracy.

It should also be noted that the results of the scalar fit by DFGS are much more compatible with our main results and with the results by CGL. One might ask whether the differences between the mean values of $\bar{b}_1$ and $\alpha_{\pi\pi}$ in the DFGS extended (and global) fit and their scalar fit, the CGL results, and our NA48/2 results are caused by the CGL correlation assumption (\ref{eq:a20}) or whether they are driven by the BNL-E865 data. Our ETM results, independent of (\ref{eq:a20}), would suggest that it is the latter, but as was previously discussed, they might be potentially plagued by a large systematic uncertainty stemming from the ETM calculation of $a_0^0$. 

Therefore, we can turn to our NA48/2+ETM refit (\ref{eq:etm_refit}) in order to dispense with the assumption (\ref{eq:a20}). In addition, we can employ the combination of ETM+CGL correlation (\ref{eq:etm_CGL_corr}) to get a result independent of both the NA48/2 and BNL-E865 data, as well as the uncertain ETM calculation of $a_0^0$. The obtained values are listed in Table \ref{table:etm_new}.

\begin{table}[H]
\centering\footnotesize
    \begin{tabular}{|c|c|c|c|c||c|c|}
        \hline \rule[-0.2cm]{0cm}{0.75cm}
        & \textcolor{red}{NA48/2} & \tcr{GKPRY CFD} & \textcolor{red}{RBC/UKQCD} & \textcolor{red}{ETM} & \bf DFGS \cite{DFGS} & \bf CGL \cite{CGL}  \\
        \hline
        \hline \rule[-0.1cm]{-0.1cm}{0.5cm}
        $\bar{b}_{1}$& $-12.30\pm 1.88$ & $-8.49\pm 6.12$&$-17.77 \pm 8.26$ &$-10.58 \pm 0.66$ & $-1.51\pm 7.01$ & $-12.4\pm 1.6$\\
        \hline \rule[-0.1cm]{-0.1cm}{0.5cm}
        $\bar{b}_{2}$& 11.21$\pm$0.67 & 9.92$\pm$3.33 & 12.72$\pm$6.78 & 6.82$\pm$2.09 & 8.93$\pm$1.62 & 11.8$\pm$0.6 \\
        \hline \rule[-0.1cm]{-0.1cm}{0.5cm}
        $\bar{b}_{3}$& $-0.26\pm 0.06$ & $-0.35\pm 0.06$ &
        $-0.23\pm 0.15$ &
        $-0.34\pm 0.06$ & $-0.36\pm 0.07$ & $-0.33\pm 0.07$ \\
        \hline \rule[-0.1cm]{-0.1cm}{0.5cm}
        $\bar{b}_{4}$& 0.73$\pm$0.01 & 0.73$\pm$0.02 &
        0.75$\pm$0.06 &
        0.69$\pm$0.02 & 0.71$\pm$0.01 & 0.74$\pm$0.01 \\
        \hline \rule[-0.1cm]{-0.1cm}{0.5cm}
        $\bar{b}_{5}$& 3.67$\pm$0.42 & 3.75$\pm$0.73 &
        4.09$\pm$1.45 &
        2.88$\pm$0.49 & 3.21$\pm$0.44 & 3.58$\pm$0.37 \\
        \hline \rule[-0.1cm]{-0.1cm}{0.5cm}
        $\bar{b}_{6}$& 2.31$\pm$0.03 & 2.27$\pm$0.16 &
        2.38$\pm$0.28 &
        2.13$\pm$0.08 & 2.2$\pm$0.08 & 2.35$\pm$0.02 \\
        \hline
        \hline \rule[-0.1cm]{-0.1cm}{0.5cm}
        $\alpha_{\pi\pi}$ & 1.053$\pm$0.071 & 1.139$\pm$0.167 &
        0.905$\pm$0.366 &0.872$\pm$0.107 & 1.384$\pm$0.267 & 1.08$\pm$0.07 \\
        \hline \rule[-0.1cm]{-0.1cm}{0.5cm}
        $\beta_{\pi\pi}$ & 1.115$\pm$0.008 & 1.092$\pm$0.047 &
        1.139$\pm$0.105 &1.048$\pm$0.032 & 1.077$\pm$0.025 & 1.12$\pm$0.01 \\
        \hline \rule[-0.1cm]{-0.1cm}{0.5cm}
        $\rho_{\alpha\beta}$ & $-0.16$ & $-0.14$ & 0.56 & 0.97 & $-0.23$ &\\
        \hline
        \hline \rule[-0.1cm]{-0.1cm}{0.5cm}
        $10^{-3}\lambda_{1}$ & $-3.51\pm 0.66$ & $-4.01\pm 0.71$ &
        $-3.18\pm 1.35$ &
        $-4.11\pm 0.63$ & $-4.18\pm 0.63$ & \\
        \hline \rule[-0.1cm]{-0.1cm}{0.5cm}
        $10^{-3}\lambda_{2}$ & 9.20$\pm$0.11 &  9.24$\pm$0.22 &
        9.43$\pm$0.81 & 
        8.65$\pm$0.26 & 8.96$\pm$0.12 & \\
        \hline \rule[-0.1cm]{-0.1cm}{0.5cm}
        $10^{-4}\lambda_{3}$ & 2.38$\pm$0.17 & 2.42$\pm$0.30 &
        2.54$\pm$0.55 &
        2.09$\pm$0.19 & 2.22$\pm$0.16 & \\
        \hline \rule[-0.1cm]{-0.1cm}{0.5cm}
        $10^{-4}\lambda_{4}$ & $-1.44\pm 0.02$ & $-1.42\pm 0.09$ &
        $-1.48\pm 0.15$ &
        $-1.35\pm 0.04$ & $-1.38\pm 0.04$ & \\
        \hline
    \end{tabular}
\caption{Comparison of \textcolor{red}{our results} with literature}
\label{table:bi}
\end{table}

\begin{table}[H]
\centering\footnotesize
    \begin{tabular}{|c|c|c|c||c|c|}
        \hline \rule[-0.2cm]{0cm}{0.75cm}
        & \tcr{NA48/2+ETM} & \tcr{ETM+CGL} & \tcr{NA48/2} & \bf \multirow{2}{1.65cm}{DFGS \cite{DFGS}} & \bf \;\multirow{2}{1.4cm}{CGL \cite{CGL}} \\
        & \tcr{refit} & \tcr{correlation} &  \tcr{($\theta_{0,1}$ no error)} & & \\
        \hline
        \hline \rule[-0.1cm]{-0.1cm}{0.5cm} $\bar{b}_{1}$& $-11.24\pm 0.58$ & $-11.30\pm 0.56$ & $-12.30\pm 1.88$ &$-1.51\pm 7.01$&$-12.4\pm 1.6$\\
        \hline \rule[-0.1cm]{-0.1cm}{0.5cm} $\bar{b}_{2}$& 10.95$\pm$0.68 & 11.31$\pm$0.76 & 11.23$\pm$0.54 & 8.93$\pm$1.62&$11.8\pm 0.6$\\
        \hline \rule[-0.1cm]{-0.1cm}{0.5cm} $\bar{b}_{3}$& $-0.27\pm 0.06$ & $-0.26\pm 0.06$ & $-0.27\pm 0.01$ & $-0.36\pm 0.07$&$-0.33\pm 0.07$\\
        \hline \rule[-0.1cm]{-0.1cm}{0.5cm} $\bar{b}_{4}$& 0.73$\pm$0.01 & 0.73$\pm$0.01 & 0.73$\pm$0.01 & 0.71$\pm$0.01& $0.74\pm 0.01$\\
        \hline \rule[-0.1cm]{-0.1cm}{0.5cm} $\bar{b}_{5}$& 3.62$\pm$0.42 & 3.69$\pm$0.43 & 3.63$\pm$0.10 & 3.21$\pm$0.44&$3.58\pm 0.37$\\
        \hline \rule[-0.1cm]{-0.1cm}{0.5cm} $\bar{b}_{6}$& 2.29$\pm$0.03 & 2.31$\pm$0.03 & 2.31$\pm$0.03 & 2.2$\pm$0.08&$2.35\pm 0.02$\\
        \hline
        \hline \rule[-0.1cm]{-0.1cm}{0.5cm} $\alpha_{\pi\pi}$ & 1.084$\pm$0.034 & 1.103$\pm$0.043 & 1.054$\pm$0.070 & 1.384$\pm$0.267 & 1.08$\pm$0.07 \\
        \hline \rule[-0.1cm]{-0.1cm}{0.5cm} $\beta_{\pi\pi}$ & 1.111$\pm$0.008 & 1.117$\pm$0.010 & 1.115$\pm$0.008 & 1.077$\pm$0.025 & 1.12$\pm$0.01 \\
        \hline \rule[-0.1cm]{-0.1cm}{0.5cm} $\rho_{\alpha\beta}$ & 0.66 & 0.80 & $-0.19$ &  $-0.23$ & \\
        \hline
        \hline \rule[-0.1cm]{-0.1cm}{0.5cm} $10^{-3}\lambda_{1}$ & $-3.55\pm 0.65$ & $-3.50\pm 0.66$ & $-3.56\pm 0.10$ & $-4.18\pm 0.63$ & \\
        \hline \rule[-0.1cm]{-0.1cm}{0.5cm} $10^{-3}\lambda_{2}$ & 9.17$\pm$0.11 & 9.21$\pm$0.12 & 9.20$\pm$0.09 & 8.96$\pm$0.12 & \\
        \hline \rule[-0.1cm]{-0.1cm}{0.5cm} $10^{-4}\lambda_{3}$ & 2.36$\pm$0.17 & 2.39$\pm$0.17 & 2.37$\pm$0.04 & 2.22$\pm$0.16 & \\
        \hline \rule[-0.1cm]{-0.1cm}{0.5cm} $10^{-4}\lambda_{4}$ & $-1.43\pm 0.02$ & $-1.44\pm 0.02$ & $-1.44\pm 0.02$ & $-1.38\pm 0.04$ & \\
        \hline
    \end{tabular}
\caption{\textcolor{red}{Our results} with alternative inputs }
\label{table:etm_new}
\end{table}

As can be seen, our results are highly compatible with each other, although we effectively removed the CGL correlation (\ref{eq:a20}) from the NA48/2+ETM refit. Both of these results, based on the ETM calculation of $a_0^2$, are in good agreement with CGL \cite{CGL}, while some tension remains with DFGS \cite{DFGS}. It is notable that the error bars are reduced compared to all the results in Table \ref{table:bi} and the low values of $\alpha_{\pi\pi}$ and $\bar{b}_1$ are confirmed with higher precision.

Table \ref{table:etm_new} also includes a check for the dependence of the results on the uncertainty in the phase shifts at the matching point $s_0$. We have performed the calculation while setting the errors of $\theta_0$ and $\theta_1$ to zero, highlighting the sensitivity of the results to these uncertainties compared to the NA48/2 values in Table \ref{table:bi}. Several of the parameters have turned out to be highly sensitive ($\overline{b}_3$, $\overline{b}_5$, $\lambda_1$ and $\lambda_3$), so improved precision for the phase-shift values could be beneficial for their determination. However, the dependence is negligible for $\alpha_{\pi\pi}$ and $\beta_{\pi\pi}$, which are the main focus of our article.

\section{Summary}

In this article, we have extracted the subthreshold parameters $\alpha_{\pi\pi}$, $\beta_{\pi\pi}$, $\lambda_i$, and the coefficients $\bar{b}_i$ of dispersive representations of the $\pi\pi$ scattering amplitude. We have used recent experimental data and lattice QCD calculations of the $\pi\pi$ scattering lengths as our inputs, while employing the solutions to the Roy equations developed in \cite{ACGL, DFGS}. We have numerically modeled the probability distribution of the results based on all uncertainties in the inputs by Monte Carlo sampling.

As our main result based on experimental data we have utilized the analysis by the NA48/2 collaboration (model C) \cite{NA48}, which obtained the scattering lengths by combining measurements of the $K_{e4}$ decay and the cusp in $K\to 3\pi$. Their Model C relied on a theoretical correlation between $a_0^0$ and $a_0^2$ (\ref{eq:a20}), which is based on the relation between the scalar pion radius and the scattering lengths \cite{CGL, CGL_korelace}, established in the framework of two-flavor $\chi$PT. We have obtained the following values for $\alpha_{\pi\pi}$ and $\beta_{\pi\pi}$:
\begin{equation}
    \hspace{0cm}
    \alpha_{\pi\pi} = 1.053(71), \quad \beta_{\pi\pi} = 1.115(8), \quad \rho_{\alpha\beta} = -0.16 \qquad \mathrm{(NA48/2\ (model\ C))}, 
\end{equation}

\no The full list of values for all parameters can be found in Table \ref{table:bi}, as well as the results for alternative inputs, as described in Sec. \ref{subsec:data}. 

We have also performed the analysis with lattice QCD inputs (see Table \ref{table:scatter_len}). Our main result (NA48/2+ETM refit) uses the combination of NA48/2 data and the calculation of the scattering length $a_0^2$ by the ETM collaboration \cite{ETM_a20}, as described in Sec. \ref{subsec:data}, which dispenses with the theoretical assumption (\ref{eq:a20}): 
\begin{equation}
    \hspace{0.25cm}
    \alpha_{\pi\pi} = 1.084(34), \quad \beta_{\pi\pi} = 1.111(8), \quad \rho_{\alpha\beta} = 0.66 \qquad\ \mathrm{(NA48/2+ETM\ refit)},
\end{equation}

\no The complete set can be found in Table \ref{table:etm_new}. 

As can be seen, our results are highly compatible with each other and with the 2-flavor $\chi$PT results by CGL \cite{CGL}, although we effectively removed the CGL correlation (\ref{eq:a20}). In addition, we can exclude high values of $\alpha_{\pi\pi}$ and $\bar{b}_1$, as suggested by the mean values of global/extended fits obtained by DFGS \cite{DFGS}, which used the BNL-E865 data without employing the CGL correlation. Hence, we conclude that the theoretical assumption (\ref{eq:a20}) is not the source of this tension. 

It should also be noted that our results for both $\alpha_{\pi\pi}$ and $\beta_{\pi\pi}$ are fairly close to unity, which is the leading order value in the framework of 3-flavor $\chi$PT, as discussed in the Introduction. Thus, it is compatible with the expectation of good convergence for these parameters. In particular, we expect the lower value of $\alpha_{\pi\pi}$, compared to the one obtained by DFGS \cite{DFGS}, to be more compatible with the $\eta\to 3\pi$ data, based on the analysis \cite{Kolesar_2018} and to not lead to a suppressed pion mass at the leading order.

We intend to utilize the obtained results in order to update the analysis \cite{Kolesar_2018}, combining $\pi\pi$ and $\eta\to 3\pi$ data to extract the value of the leading-order 3-flavor $\chi$PT LEC $B_0$. Similarly, our analysis \cite{Kolesar_2023}, which focused on NLO LECs $L_4$ and $L_5$, can be extended by the inclusion of $\pi\pi$ scattering lengths in a relatively straightforward way. \\

\no \textbf{Acknowledgments:} The work presented in this paper was supported by the Czech Science Foundation Grant No. 23-06770S and by the Charles University Grant No. PRIMUS 23/SCI/025. 

\bibliographystyle{elsarticle-mod}
\bibliography{bibliography}

\newpage
\appendix
\section{Unitarity corrections $\overline{K}(s|t,u)$} \label{Kstu}
\begin{align}
\overline{K}(s|t,u) = &32\pi\sum_{n=0}^{4}(\frac{1}{3}\left[w_{0}^{(n)}(s)-w_{2}^{(n)}(s)\right]\overline{K}_{n}(s)+\frac{1}{2}\left[w_{2}^{(n)}(t)+3(s-u)w_{1}^{(n)}(t)\right]\overline{K}_{n}(t)+\\
\nonumber &+\frac{1}{2}\left[w_{2}^{(n)}(u)+3(s-t)w_{1}^{(n)}(u)\right]\overline{K}_{n}(u)),
\end{align}
where 
\begin{align}
\overline{K}_{0}(s) = &\overline{J}(s)\\
\overline{K}_{1}(s) = &\frac{1}{16\pi^{2}}\frac{s}{s-4M_{\pi}^{2}}\left(16\pi^{2}\overline{J}(s)-2\right)^{2}\\
\overline{K}_{2}(s) = &\frac{s-4M_{\pi}^{2}}{s}\overline{K}_{1}(s)-\frac{1}{4\pi^{2}}\\
\overline{K}_{3}(s) = &\frac{1}{16\pi^{2}}\frac{M_{\pi}^{2}}{s-4M_{\pi}^{2}}\left(\frac{s}{s-4M_{\pi}^{2}}\left(16\pi^{2}\overline{J}(s)-2\right)^{3}+\pi^{2}\left(16\pi^{2}\overline{J}(s)-2\right)\right)-\frac{1}{32}\\
\overline{K}_{4}(s) = &\frac{1}{16\pi^{2}}\frac{M_{\pi}^{2}}{s-4M_{\pi}^{2}}\left(16\pi^{2}\overline{J}(s)-2+8\pi^{2}\overline{K}_{1}(s)+\frac{16\pi^{2}}{3}\overline{K}_{3}(s)+\frac{\pi^{2}}{3}\right)-\frac{1}{32\pi^{2}}+\frac{1}{192}\\
\overline{J}(s) = &\frac{s}{16\pi^{2}}\int_{4M_{\pi}^{2}}^{\infty}\frac{dx}{x}\frac{1}{x-s}\sqrt{\frac{x-4M_{\pi}^{2}}{x}}\\
w_{a}^{(n)}(s) = &16\pi|\phi_{a}(s)|^{2}\delta_{n0}+\frac{2M_{\pi}^{4}}{F_{\pi}^{4}}\phi_{a}(s)\xi_{a}^{(n)}(s)\\
\phi_{0}(s) = &\frac{1}{96\pi F_{\pi}^{2}}\left(6\beta_{\pi\pi}(s-\frac{4}{3}M_{\pi}^{2})+5\alpha_{\pi\pi}M_{\pi}^{2}\right)\\
\phi_{1}(s) = &\frac{1}{96\pi F_{\pi}^{2}}\beta_{\pi\pi}(s-4M_{\pi}^{2})\\
\phi_{2}(s) = &\frac{1}{96\pi F_{\pi}^{2}}\left(-3\beta_{\pi\pi}(s-\frac{4}{3}M_{\pi}^{2})+2\alpha_{\pi\pi}^{2}\right).
\end{align}

We will define
\begin{align}
q=\sqrt{\frac{s-4M_{\pi}^{2}}{4M_{\pi}^{2}}},
\end{align}
and thus
\begin{align}
\nonumber \xi_{0}^{(0)}(s)=&\frac{1}{144\pi^{2}}(35\alpha_{\pi\pi}^{2}+80\alpha_{\pi\pi}\beta_{\pi\pi}+134\beta_{\pi\pi}^{2})+10(\lambda_{1}+2\lambda_{2})+\\
&+\left(\frac{1}{72\pi^{2}}(60\alpha_{\pi\pi}+209\beta_{\pi\pi})\beta_{\pi\pi}+16(2\lambda_{1}+3\lambda_{2})\right)q^{2}+\\
\nonumber &+\left(\frac{311}{108\pi^{2}}\beta_{\pi\pi}^{2}+\frac{8}{3}(11\lambda_{1}+14\lambda_{2})\right)q^{4}\\
\xi_{0}^{(1)}(s)=&\frac{1}{192\pi^{2}}\left(5\alpha_{\pi\pi}^{2}+4\beta_{\pi\pi}^{2}\right)+\frac{1}{9\pi^{2}}\beta_{\pi\pi}q^{2}+\frac{7}{36\pi^{2}}\beta_{\pi\pi}^{2}q^{4}\\
\xi_{0}^{(2)}(s)=&\frac{1}{1152\pi^{2}}\left(5\alpha_{\pi\pi}+16\beta_{\pi\pi}+24\beta_{\pi\pi}q^{2}\right)^{2}\\
\xi_{0}^{(3)}(s)=&\frac{1}{288\pi^{2}}\left(-5\alpha_{\pi\pi}^{2}+4\beta_{\pi\pi}^{2}\right)+\frac{1}{12\pi^{2}}\beta_{\pi\pi}^{2}q^{2}\\
\xi_{0}^{(4)}(s)=&0\\
\nonumber \xi_{2}^{(0)}(s)=&\frac{1}{288\pi^{2}}\left(31\alpha_{\pi\pi}^{2}-122\alpha_{\pi\pi}\beta_{\pi\pi}+220\beta_{\pi\pi}^{2}\right)+4(\lambda_{1}+2\lambda_{2})+\\
&+\left(\frac{1}{144\pi^{2}}(-69\alpha_{\pi\pi}+268\beta_{\pi\pi})\beta_{\pi\pi}+8(\lambda_{1}+3\lambda_{2})\right)q^{2}+\\
\nonumber &\left(\frac{265}{216\pi^{2}}\beta_{\pi\pi}^{2}+\frac{16}{3}(\lambda_{1}+4\lambda_{2})\right)q^{4}\\
\nonumber \xi_{2}^{(1)}(s)=&\frac{1}{576\pi^{2}}\left(9\alpha_{\pi\pi}^{2}-42\alpha_{\pi\pi}\beta_{\pi\pi}+60\beta_{\pi\pi}^{2}\right)+\\
&+\frac{1}{144\pi^{2}}(-9\alpha_{\pi\pi}+37\beta_{\pi\pi})\beta_{\pi\pi}q^{2}+\\
\nonumber &+\frac{11}{72\pi^{2}}\beta_{\pi\pi}^{2}q^{4}\\
\xi_{2}^{(2)}(s)=&\frac{1}{288\pi^{2}}\left(\alpha_{\pi\pi}-4\beta_{\pi\pi}-6\beta_{\pi\pi}q^{2}\right)^{2}\\
\xi_{2}^{(3)}(s)=&\frac{1}{288\pi^{2}}\left(-3\alpha_{\pi\pi}^{2}+2\alpha_{\pi\pi}\beta_{\pi\pi}-12\beta_{\pi\pi}^{2}\right)-\frac{1}{24\pi^{2}}\beta_{\pi\pi}^{2}q^{2}\\
\xi_{2}^{(4)}(s)=&0\\
\nonumber\xi_{1}^{(0)}(s)=&\frac{1}{576\pi^{2}}\left(5\alpha_{\pi\pi}^{2}-80\alpha_{\pi\pi}\beta_{\pi\pi}+10\beta_{\pi\pi}^{2}\right)+\\
&+\left(\frac{1}{432\pi^{2}}\left(55\alpha_{\pi\pi}-68\beta_{\pi\pi}\right)\beta_{\pi\pi}-\frac{8}{3}(\lambda_{1}-\lambda_{2})\right)q^{2}-\\
\nonumber&-\left(\frac{\beta_{\pi\pi}^{2}}{108\pi^{2}}+\frac{8}{3}(\lambda_{1}-\lambda_{2})\right)q^{4}\\
\xi_{1}^{(1)}(s)=&\frac{1}{288\pi^{2}}(-5\alpha_{\pi\pi}+7\beta_{\pi\pi})\beta_{\pi\pi}+\frac{1}{144\pi^{2}}(5\alpha_{\pi\pi}-3\beta_{\pi\pi})\beta_{\pi\pi}q^{2}-\frac{1}{72\pi^{2}}\beta_{\pi\pi}^{2}q^{4}\\
\xi_{1}^{(2)}(s)=&\frac{1}{72\pi^{2}}\beta_{\pi\pi}^{2}q^{4}\\
\xi_{1}^{(3)}(s)=&\frac{1}{864\pi^{2}}\left(-5\alpha_{\pi\pi}^{2}+10\alpha_{\pi\pi}\beta_{\pi\pi}+28\beta_{\pi\pi}^{2}\right)+\frac{1}{24\pi^{2}}\beta_{\pi\pi}^{2}q^{2}\\
\xi_{1}^{(4)}(s)=&-\frac{5}{144\pi^{2}}\left(\alpha_{\pi\pi}^{2}+4\alpha_{\pi\pi}\beta_{\pi\pi}-2\beta_{\pi\pi}^{2}\right)
\end{align}

\newpage
\section{Solutions of Roy equations} \label{schenkpar}
In ACGL \cite{ACGL}, the authors have calculated the solutions of Roy equations. These solutions consider the three lowest partial-wave amplitudes, valid in the range $2M_{\pi}\leq\sqrt{s}\leq0.8 \mathrm{GeV}\equiv \sqrt{s_0}$. As was mentioned in Section \ref{Sec:pipi_scatt}, the amplitudes are expressed as
\begin{eqnarray}
t_{l}^{I}(s)=\frac{1}{\sigma(s)}e^{i\delta_{l}^{I}(s)}\sin(\delta_{l}^{I}(s))
\end{eqnarray}
and use the parametrization proposed by Schenk \cite{SCHENK}
\begin{eqnarray}
\tan(\delta_{l}^{I})=\sqrt{1-\frac{4M_{\pi}^{2}}{s}}q^{2l}\left(A_{l}^{I}+B_{l}^{I}q^{2}+C_{l}^{I}q^{4}+D_{l}^{I}q^{6}\right)\frac{4M_{\pi}^{2}-s_{l}^{I}}{s-s_{l}^{I}},
\end{eqnarray}
where $q$ is
\begin{equation}
q=\frac{1}{2}\sqrt{s-4M_{\pi}^2}.
\end{equation}
The Schenk parameters can be approximated by a third degree polynomial in scattering lengths, using $A_{0}^{0}$ as an example
\begin{equation}
A_{0}^{0}=z_{1}+z_{2}v+z_{4}u^{2}+z_{5}v^{2}+z_{6}uv+z_{7}u^{3}+z_{8}u^{2}v+z_{9}uv^{2}+z_{10}v^{3}.
\end{equation}
Here \emph{u} and \emph{v} are
\begin{align}
u=\frac{a_{0}^{0}}{p_{0}}-1, v=\frac{a_{0}^{2}}{p_{2}}-1\\
p_{0}=0.225, p_{2}=-0.03706,
\end{align}
where the $z_{i}$'s are the coefficients shown in Table \ref{roysol}, adopted from ACGL \cite{ACGL}, in $M_{\pi}^{2n}$ units.

\begin{table}
    \centering \small
    \begin{tabular}{|c||c|c|c|c|c|}
    \hline
         & $A_0^0$ & $B_0^0$ & $C_0^0$ & $D_0^0$ & $s_0^0$\\
         \hline\hline
         $z_1$&0.225 &0.2463 &$-0.1665\cdot 10^{-1}$ &$-0.6403\cdot 10^{-3}$ &$0.3672\cdot 10^{2}$\\
         \hline
         $z_2$&0.225 &0.1985 &$0.3283\cdot 10^{-2}$ &$-0.4136\cdot 10^{-2}$ &$0.1339\cdot 10$\\
         \hline
         $z_3$&0 &0.1289 &$0.1142\cdot 10^{-1}$ &$-0.3699\cdot 10^{-2}$ &0.6504\\
        \hline
         $z_4$&0 &$0.1426\cdot 10^{-1}$ &$0.14\cdot 10^{-1}$ &$-0.398\cdot 10^{-2}$ &$-0.3211\cdot 10$\\
         \hline
         $z_5$&0 &$0.8717\cdot 10^{-2}$ &$0.1613\cdot 10^{-1}$ &$-0.3152\cdot 10^{-2}$ &$-0.1396\cdot 10$\\
         \hline
         $z_6$&0 &$0.5058\cdot 10^{-1}$ &$0.3\cdot 10^{-1}$ &$-0.7354\cdot 10^{-2}$ &$-0.4114\cdot 10$\\
         \hline
         $z_7$&0 &$-0.4266\cdot 10^{-2}$ &$-0.4045\cdot 10^{-2}$ &$-0.1212\cdot 10^{-2}$ &$-0.3447\cdot 10$\\
         \hline
         $z_8$&0 &$-0.4658\cdot 10^{-2}$ &$0.211\cdot 10^{-2}$ &$-0.4544\cdot 10^{-2}$ &$-0.8428\cdot 10$\\
         \hline
         $z_9$&0 &$-0.5358\cdot 10^{-2}$ &$0.1095\cdot 10^{-1}$ &$-0.4558\cdot 10^{-2}$ &$-0.635\cdot 10$\\
         \hline
         $z_{10}$&0 &$-0.2555\cdot 10^{-2}$ &$0.4249\cdot 10^{-2}$ &$-0.1271\cdot 10^{-2}$ &$-0.1486\cdot 10$\\
         \hline
%    \end{tabular}\\
%\end{table}
%\begin{table}[H]
%    \centering \scriptsize
%    \begin{tabular}{|c||c|c|c|c|c|}
    \hline
         & $A_1^1$ & $B_1^1$ & $C_1^1$ & $D_1^1$ & $s_1^1$\\
         \hline\hline
         $z_1$&$0.3626\cdot 10^{-1}$ &$0.1337\cdot 10^{-3}$ &$-0.6976\cdot 10^{-4}$ &$0.1408\cdot 10^{-5}$ &$0.3074\cdot 10^{2}$\\
         \hline
         $z_2$&$0.1834\cdot 10^{-1}$ &$-0.2336\cdot 10^{-2}$ &$0.1965\cdot 10^{-3}$ &$-0.1974\cdot 10^{-4}$ &$-0.2459$\\
         \hline
         $z_3$&$0.1081\cdot 10^{-1}$ &$-0.8563\cdot 10^{-3}$ &$0.3268\cdot 10^{-4}$ &$-0.8821\cdot 10^{-5}$ &$-0.1733$\\
        \hline
         $z_4$&$-0.3195\cdot 10^{-2}$ &$0.1678\cdot 10^{-3}$ &$0.2173\cdot 10^{-4}$ &$-0.6047\cdot 10^{-6}$ &$0.6323\cdot 10^{-1}$\\
         \hline
         $z_5$&$0.167\cdot 10^{-3}$ &$0.4147\cdot 10^{-4}$ &$0.3267\cdot 10^{-5}$ &$-0.1617\cdot 10^{-5}$ &$-0.109\cdot 10^{-2}$\\
         \hline
         $z_6$&$-0.9543\cdot 10^{-3}$ &$0.8402\cdot 10^{-4}$ &$0.2059\cdot 10^{-4}$ &$-0.3125\cdot 10^{-5}$ &$0.2724\cdot 10^{-1}$\\
         \hline
         $z_7$&$0.5049\cdot 10^{-3}$ &$-0.9308\cdot 10^{-4}$ &$0.107\cdot 10^{-4}$ &$-0.1257\cdot 10^{-5}$ &$-0.7218\cdot 10^{-2}$\\
         \hline
         $z_8$&$0.4595\cdot 10^{-4}$ &$-0.2755\cdot 10^{-3}$ &$0.5554\cdot 10^{-4}$ &$-0.4432\cdot 10^{-5}$ &$0.1483\cdot 10^{-1}$\\
         \hline
         $z_9$&$-0.9\cdot 10^{-4}$ &$-0.2308\cdot 10^{-3}$ &$0.5307\cdot 10^{-4}$ &$-0.4415\cdot 10^{-5}$ &$0.1813\cdot 10^{-1}$\\
         \hline
         $z_{10}$&$-0.1198\cdot 10^{-4}$ &$-0.612\cdot 10^{-4}$ &$0.1519\cdot 10^{-4}$ &$-0.1344\cdot 10^{-5}$ &$0.5016\cdot 10^{-2}$\\
         \hline
%    \end{tabular}\\
%    \begin{tabular}{|c||c|c|c|c|c|}
    \hline
         & $A_0^2$ & $B_0^2$ & $C_0^2$ & $D_0^2$ & $s_0^2$\\
         \hline\hline
         $z_1$&$-0.3706\cdot 10^{-1}$ &$-0.8553\cdot 10^{-1}$ &$-0.7542\cdot 10^{-2}$ &$0.1987\cdot 10^{-3}$ &$-0.1192\cdot 10^{2}$\\
         \hline
         $z_2$&0 &$-0.1236\cdot 10^{-1}$ &$0.3466\cdot 10^{-1}$ &$-0.2524\cdot 10^{-2}$ &$-0.404\cdot 10^{2}$\\
         \hline
         $z_3$&$-0.3706\cdot 10^{-1}$ &$-0.6673\cdot 10^{-2}$ &$0.2857\cdot 10^{-1}$ &$-0.1993\cdot 10^{-2}$ &$-0.3457\cdot 10^{2}$\\
        \hline
         $z_4$&0 &$0.4901\cdot 10^{-2}$ &$0.2674\cdot 10^{-2}$ &$0.1506\cdot 10^{-2}$ &$-0.9879\cdot 10^{2}$\\
         \hline
         $z_5$&0 &$0.281\cdot 10^{-1}$ &$0.1477\cdot 10^{-1}$ &$0.2915\cdot 10^{-3}$ &$-0.9856\cdot 10^{2}$\\
         \hline
         $z_6$&0 &$0.401\cdot 10^{-1}$ &$0.2458\cdot 10^{-1}$ &$0.1325\cdot 10^{-2}$ &$-0.2072\cdot 10^{3}$\\
         \hline
         $z_7$&0 &$-0.1663\cdot 10^{-1}$ &$-0.303\cdot 10^{-1}$ &$0.8759\cdot 10^{-3}$ &$-0.1589\cdot 10^{3}$\\
         \hline
         $z_8$&0 &$-0.6784\cdot 10^{-1}$ &$-0.9512\cdot 10^{-1}$ &$0.4713\cdot 10^{-2}$ &$-0.5259\cdot 10^{3}$\\
         \hline
         $z_9$&0 &$-0.5429\cdot 10^{-1}$ &$-0.8744\cdot 10^{-1}$ &$0.5313\cdot 10^{-2}$ &$-0.5366\cdot 10^{3}$\\
         \hline
         $z_{10}$&0 &$-0.1178\cdot 10^{-1}$ &$-0.2535\cdot 10^{-1}$ &$0.173\cdot 10^{-2}$ &$-0.1723\cdot 10^{3}$\\
         \hline
    \end{tabular}
    \caption{Solutions the Schenk parametrization by ACGL \cite{ACGL}}
    \label{roysol}
\end{table}

In this article we also use the parametrization from DFGS \cite{DFGS}, which accounts for uncertainties in phase shift measurements at the matching point $\sqrt{s_0}=0.8$ GeV. It does so by assuming 
\begin{equation}
    z_j = a_j + \delta\theta_0 b_j + \delta\theta_1 c_j,
\end{equation}

\no where $\delta\theta_i$ are errors of the phase shifts at the boundary $s_0$. In this parametrization, parameters $s^0_0$, $s^1_1$ and $s^2_0$ are fixed by boundary conditions

\begin{align}
    \nonumber \delta^0_0(s_0) &\equiv \theta_0 \hspace{2.05cm} \delta^1_1(s_0) \equiv \theta_1 \hspace{2cm} \delta^2_0(s_0) \equiv \theta_2\\
    \delta\theta_{0}&=\theta_{0}-82.3\degree \hspace{1.125cm} \delta\theta_{1}=\theta_{1}-108.9\degree\\
    \nonumber \theta_0&=82.3\degree \pm 3.4\degree \hspace{1cm}\theta_1=110.4\degree \pm 0.7\degree
\end{align}

\noindent and $\theta_2(a^0_0, a^2_0, \theta_0,\theta_1)$ is also given by Schenk parametrization. Here we use the improved determination for $\theta_1$ by \cite{Colangelo:2018mtw}. The DFGS \cite{DFGS} results for $z_j$'s are
\begin{align}
    A_0^0 &: a_1 = a_2 = 0.225\\
    A^2_0 &: a_1 = a_3 = -0.03706
\end{align}

\no and the rest can be found in Table \ref{table:app_schenk_descotes}.

\begin{table}[H]
    \centering
    \scriptsize
    \begin{tabular}{|c|c||c|c|c|}
        \hline
         & $z_i$ & $a_i$ & $b_i$ & $c_i$ \\
         \hline\hline
         \multirow{10}{*}{$A^1_1$}& 1 & $0.3617\cdot 10^{-1}$ & $-0.1713\cdot 10^{-2}$& $-0.3860\cdot 10^{-2}$ \\
         \cline{2-5}
         & 2 & $0.1574\cdot 10^{-1}$ & $-0.2448\cdot 10^{-2}$& $-0.3384\cdot 10^{-3}$\\
         \cline{2-5}
         & 3 & $0.1057 \cdot 10^{-1}$ & $-0.1774 \cdot 10^{-2}$& $-0.2510 \cdot 10^{-4}$ \\
         \cline{2-5}
         & 4 & $-0.1782 \cdot 10^{-2}$ & $-0.1025 \cdot 10^{-1}$& $-0.4312 \cdot 10^{-2}$ \\
         \cline{2-5}
         & 5 & $0.2572 \cdot 10^{-3}$ & $-0.4649 \cdot 10^{-2}$ &  $-0.1705 \cdot 10^{-2}$\\
         \cline{2-5}
         & 6 & $-0.2872 \cdot 10^{-3}$ & $0.1046 \cdot 10^{-2} $ & $-0.3467 \cdot 10^{-2}$ \\
         \cline{2-5}
         & 7 & $0.8311 \cdot 10^{-2}$ & $-0.9152 \cdot 10^{-2}$ & $-0.3637 \cdot 10^{-2}$ \\
         \cline{2-5}
         & 8 & $-0.2603 \cdot 10^{-2}$ & $-0.1489 \cdot 10^{-1}$ & $0.2188 \cdot 10^{-2}$ \\
         \cline{2-5}
         & 9 & $0.1247 \cdot 10^{-2}$ & $ 0.7639 \cdot 10^{-3}$ & $-0.1340 \cdot 10^{-2}$ \\
         \cline{2-5}
         & 10 & $-0.1186 \cdot 10^{-3}$ & $ 0.4371 \cdot 10^{-2}$ & $0.1128 \cdot 10^{-4}$ \\
         \hline
    \end{tabular}\begin{tabular}{|c|c||c|c|c|}
        \hline
         & $z_i$ & $a_i$ & $b_i$ & $c_i$ \\
         \hline\hline
         \multirow{10}{*}{$B^0_0$}& 1 & $0.2482$ & $0.4902 \cdot 10^{-1} $ & $ 0.1282 \cdot 10^{-1}$ \\
         \cline{2-5}
         & 2 & $0.1997$ & $0.1630$ & $-0.3179 \cdot 10^{-3}$ \\
         \cline{2-5}
         & 3 & $0.1285$ & $0.1137$ & $0.1640 \cdot 10^{-3}$ \\
         \cline{2-5}
         & 4 & $0.1831 \cdot 10^{-1}$ & $-0.1185$ & $0.6305 \cdot 10^{-1}$ \\
         \cline{2-5}
         & 5 & $0.9970 \cdot 10^{-2}$ & $-0.6395 \cdot 10^{-2}$ & $0.1104 \cdot 10^{-1}$ \\
         \cline{2-5}
         & 6 & $0.4846 \cdot 10^{-1}$ & $0.3431$ & $-0.1661 \cdot 10^{-1}$  \\
         \cline{2-5}
         & 7 & $-0.3888 \cdot 10^{-2}$ & $-0.1598$ & $0.4322 \cdot 10^{-1}$ \\
         \cline{2-5}
         & 8 & $-.8912 \cdot 10^{-2}$ & $0.5183$ & $-0.3067 \cdot 10^{-1}$ \\
         \cline{2-5}
         & 9 & $-0.4265 \cdot 10^{-2}$ & $0.4161 \cdot 10^{-1}$ & $0.8623 \cdot 10^{-2}$ \\
         \cline{2-5}
         & 10 & $-0.3232 \cdot 10^{-2}$ & $-0.1073$ & $0.2976 \cdot 10^{-2}$  \\
         \hline
    \end{tabular}\\
    \begin{tabular}{|c|c||c|c|c|}
        \hline
         & $z_i$ & $a_i$ & $b_i$ & $c_i$ \\
         \hline\hline
         \multirow{10}{*}{$B^1_1$}& 1 & $0.1135 \cdot 10^{-3}$ & $-0.1685 \cdot 10^{-3}$ & $-0.6043 \cdot 10^{-3}$ \\
         \cline{2-5}
         & 2 & $-0.2094 \cdot 10^{-2}$ & $-0.3429 \cdot 10^{-3}$ & $-0.5583 \cdot 10^{-4}$ \\
         \cline{2-5}
         & 3 & $-0.8626 \cdot 10^{-3}$ & $-0.2467 \cdot 10^{-3}$ & $-0.2205 \cdot 10^{-4}$ \\
         \cline{2-5}
         & 4 & $0.2911 \cdot 10^{-3}$ & $-0.8897 \cdot 10^{-3}$ & $-0.5793 \cdot 10^{-3}$ \\
         \cline{2-5}
         & 5 & $-0.5793 \cdot 10^{-3}$ & $-0.4099 \cdot 10^{-3}$ & $-0.2258 \cdot 10^{-3}$ \\
         \cline{2-5}
         & 6 & $0.2063 \cdot 10^{-3}$ & $-0.4832 \cdot 10{-3}$ & $-0.6376 \cdot 10^{-3}$ \\
         \cline{2-5}
         & 7 & $0.5294 \cdot 10^{-3}$ & $-0.6346 \cdot 10^{-3}$ & $-0.3879 \cdot 10^{-3}$ \\
         \cline{2-5}
         & 8 & $-0.3372 \cdot 10^{-3}$ & $-0.2347 \cdot 10^{-2}$ & $0.9292 \cdot 10^{-5}$ \\
         \cline{2-5}
         & 9 & $-0.1564 \cdot 10^{-3}$ & $0.1032 \cdot 10^{-4}$ & $-0.1169 \cdot 10^{-4}$ \\
         \cline{2-5}
         & 10 & $-0.1301 \cdot 10^{-4}$ & $0.8137 \cdot 10^{-3}$ & $-0.1051 \cdot 10^{-3}$ \\
         \hline
    \end{tabular}\begin{tabular}{|c|c||c|c|c|}
        \hline
         & $z_i$ & $a_i$ & $b_i$ & $c_i$ \\
         \hline\hline
         \multirow{10}{*}{$B^2_0$}& 1 & $-0.8567 \cdot 10^{-1} $ & $-0.5496 \cdot 10^{-2}$ & $0.1526 \cdot 10^{-2}$ \\
         \cline{2-5}
         & 2 & $-0.1561 \cdot 10^{-1}$ & $0.1510 \cdot 10^{-2}$ & $-0.6254 \cdot 10^{-3}$ \\
         \cline{2-5}
         & 3 & $-0.8722 \cdot 10^{-2}$ & $0.9679 \cdot 10^{-3}$ & $0.2538 \cdot 10{-3}$ \\
         \cline{2-5}
         & 4 & $0.9872 \cdot 10^{-2}$ & $0.1001 \cdot 10^{-1}$ & $0.2140 \cdot 10^{-1}$ \\
         \cline{2-5}
         & 5 & $0.2176 \cdot 10^{-1}$ & $0.3724 \cdot 10^{-2}$ & $0.3595 \cdot 10^{-2}$ \\
         \cline{2-5}
         & 6 & $0.3338 \cdot 10^{-1}$ & $-0.1050\cdot 10^{-1}$ & $-0.5945\cdot 10^{-2}$ \\
         \cline{2-5}
         & 7 & $-0.2051\cdot 10^{-1}$ & $0.4012\cdot 10^{-1}$ & $0.1157\cdot 10^{-1}$ \\
         \cline{2-5}
         & 8 & $-0.5171\cdot 10^{-1}$ & $0.7078\cdot 10^{-2}$ & $0.1593\cdot 10^{-2}$ \\
         \cline{2-5}
         & 9 & $-0.5929\cdot 10^{-1}$ & $-0.6046\cdot 10^{-2}$ & $0.1382\cdot 10^{-2}$ \\
         \cline{2-5}
         & 10 & $-0.2247\cdot 10^{-1}$& $0.4017\cdot 10^{-2}$ & $-0.1490\cdot 10^{-2}$ \\
         \hline
    \end{tabular}\\
    \begin{tabular}{|c|c||c|c|c|}
        \hline
         & $z_i$ & $a_i$ & $b_i$ & $c_i$ \\
         \hline\hline
         \multirow{10}{*}{$C^0_0$}& 1 & $-0.1652\cdot 10^{-1}$ & $0.2246\cdot 10^{-1}$ & $0.3320\cdot 10^{-2}$ \\
         \cline{2-5}
         & 2 & $0.3280\cdot 10^{-2}$ & $0.5387\cdot 10^{-1}$ & $0.9391\cdot 10^{-4}$ \\
         \cline{2-5}
         & 3 & $0.1127\cdot 10^{-1}$ & $0.2911\cdot 10^{-1}$ & $0.2303\cdot 10^{-3}$ \\
         \cline{2-5}
         & 4 & $0.1367\cdot 10^{-1}$ & $0.1198 $ & $0.9361\cdot 10^{-2}$ \\
         \cline{2-5}
         & 5 & $0.1606\cdot 10^{-1}$& $0.5107\cdot 10^{-1}$ & $0.1440\cdot 10^{-3}$ \\
         \cline{2-5}
         & 6 & $0.2990 \cdot 10^{-1}$& $-0.1170 \cdot 10^{-1}$ & $0.1345 \cdot 10^{-2}$ \\
         \cline{2-5}
         & 7 & $-0.5982 \cdot 10^{-2}$& $0.9021 \cdot 10^{-1}$ & $0.1428 \cdot 10^{-1}$  \\
         \cline{2-5}
         & 8 & $0.1923 \cdot 10^{-2}$ & $0.9601 \cdot 10^{-1}$ & $-0.4036 \cdot 10^{-2}$ \\
         \cline{2-5}
         & 9 & $0.1106 \cdot 10^{-1}$& $0.2148 \cdot 10^{-1}$ & $-0.1501 \cdot 10^{-2}$ \\
         \cline{2-5}
         & 10 & $0.3809 \cdot 10^{-2}$& $-0.2854 \cdot 10^{-1}$ & $0.2780 \cdot 10^{-2}$ \\
         \hline
    \end{tabular}\begin{tabular}{|c|c||c|c|c|}
        \hline
         & $z_i$ & $a_i$ & $b_i$ & $c_i$ \\
         \hline\hline
         \multirow{10}{*}{$C^1_1$}& 1 & $-0.7257 \cdot 10^{-4}$ & $-0.1076 \cdot 10^{-4}$ & $-0.8750 \cdot 10^{-4}$  \\
         \cline{2-5}
         & 2 & $0.2234 \cdot 10^{-3}$& $-0.4577 \cdot 10^{-4}$ & $-0.8053 \cdot 10^{-5}$  \\
         \cline{2-5}
         & 3 & $0.3718 \cdot 10^{-4}$& $-0.3531 \cdot 10^{-4}$ & $-0.6497 \cdot 10^{-5}$ \\
         \cline{2-5}
         & 4 & $0.2259 \cdot 10^{-4}$ & $0.2031 \cdot 10^{-4}$ & $-0.7306 \cdot 10^{-4}$ \\
         \cline{2-5}
         & 5 & $0.1216 \cdot 10^{-4}$ & $-0.2042 \cdot 10^{-4}$ & $-0.2856 \cdot 10^{-4}$ \\
         \cline{2-5}
         & 6 & $0.4075 \cdot 10^{-4}$& $-0.1625 \cdot 10^{-3}$ & $-0.1121 \cdot 10^{-3}$ \\
         \cline{2-5}
         & 7 & $-0.1238 \cdot 10^{-4}$ & $-0.3676 \cdot 10^{-4}$ & $-0.2568 \cdot 10^{-4}$ \\
         \cline{2-5}
         & 8 & $0.1103 \cdot 10^{-3}$& $-0.3679 \cdot 10^{-3}$ & $-0.5010 \cdot 10^{-4}$ \\
         \cline{2-5}
         & 9 & $0.3813 \cdot 10^{-4}$ & $-0.5706 \cdot 10^{-5}$ & $0.3202 \cdot 10^{-4}$ \\
         \cline{2-5}
         & 10 & $0.3531 \cdot 10^{-4}$ & $0.1373 \cdot 10^{-3}$ & $-0.3439 \cdot 10^{-4}$ \\
         \hline
    \end{tabular}\\
    \begin{tabular}{|c|c||c|c|c|}
        \hline
         & $z_i$ & $a_i$ & $b_i$ & $c_i$ \\
         \hline\hline
         \multirow{10}{*}{$C^2_0$}& 1 & $-0.7557 \cdot 10^{-2}$ & $0.2648 \cdot 10^{-2}$ & $-0.5166 \cdot 10^{-3}$ \\
         \cline{2-5}
         & 2 & $0.3425 \cdot 10^{-1}$ & $-0.2038 \cdot 10^{-2}$ & $0.5412 \cdot 10^{-3}$ \\
         \cline{2-5}
         & 3 & $0.2830 \cdot 10^{-1}$& $-0.9686 \cdot 10^{-3}$ & $0.2995 \cdot 10^{-3}$  \\
         \cline{2-5}
         & 4 & $0.3342 \cdot 10^{-2}$& $0.5536 \cdot 10^{-2}$ & $-0.5538 \cdot 10^{-2}$ \\
         \cline{2-5}
         & 5 & $0.1391 \cdot 10^{-1}$& $0.7956 \cdot 10^{-3}$ & $-0.2012 \cdot 10^{-2}$ \\
         \cline{2-5}
         & 6 & $0.2375 \cdot 10^{-1}$& $0.1775 \cdot 10^{-2}$ & $0.2675 \cdot 10^{-2}$ \\
         \cline{2-5}
         & 7 & $-0.3024 \cdot 10^{-1}$ & $-0.1924 \cdot 10^{-1}$ & $0.6680 \cdot 10^{-4}$ \\
         \cline{2-5}
         & 8 & $-0.9323 \cdot 10^{-1}$& $-0.2108 \cdot 10^{-2}$ & $0.2173 \cdot 10^{-2}$ \\
         \cline{2-5}
         & 9 & $-0.8813 \cdot 10^{-1}$ & $0.4251 \cdot 10^{-2}$ & $-0.2462 \cdot 10^{-2}$ \\
         \cline{2-5}
         & 10 & $-0.2679 \cdot 10^{-1}$& $-0.3504 \cdot 10^{-2}$ & $0.1984 \cdot 10^{-2}$ \\
         \hline
    \end{tabular}\begin{tabular}{|c|c||c|c|c|}
        \hline
         & $z_i$ & $a_i$ & $b_i$ & $c_i$ \\
         \hline\hline
         \multirow{10}{*}{$D^0_0$}& 1 & $-0.6396 \cdot 10^{-3}$ & $0.7978 \cdot 10^{-3}$ & $0.6667 \cdot 10^{-3}$  \\
         \cline{2-5}
         & 2 & $-0.4143 \cdot 10^{-2}$& $0.5649 \cdot 10^{-2}$ & $-0.5508 \cdot 10^{-4}$  \\
         \cline{2-5}
         & 3 & $-0.3708 \cdot 10^{-2}$& $0.5227 \cdot 10^{-2}$ & $0.1462 \cdot 10^{-3}$  \\
         \cline{2-5}
         & 4 & $-0.4016 \cdot 10^{-2}$& $-0.6414 \cdot 10^{-2}$ & $-0.8673 \cdot 10^{-3}$ \\
         \cline{2-5}
         & 5 & $-0.3159 \cdot 10^{-2}$& $-0.3022 \cdot 10^{-2}$ & $-0.9427 \cdot 10^{-3}$  \\
         \cline{2-5}
         & 6 & $-0.7352 \cdot 10^{-2}$ & $0.1584 \cdot 10^{-1}$ & $0.2274 \cdot 10^{-2}$  \\
         \cline{2-5}
         & 7 & $-0.1305 \cdot 10^{-2}$& $-0.1363 \cdot 10^{-1}$ & $0.3488 \cdot 10^{-2}$ \\
         \cline{2-5}
         & 8 & $-0.4523 \cdot 10^{-2}$& $0.1960 \cdot 10^{-1}$ & $0.1146 \cdot 10^{-2}$  \\
         \cline{2-5}
         & 9 & $-0.4581 \cdot 10^{-2}$& $-0.2917 \cdot 10^{-3}$ & $-0.1778 \cdot 10^{-2}$  \\
         \cline{2-5}
         & 10 & $-0.1272 \cdot 10^{-2}$& $-0.4082 \cdot 10^{-2}$ & $0.1184 \cdot 10^{-2}$  \\
         \hline
    \end{tabular}\\
    \begin{tabular}{|c|c||c|c|c|}
        \hline
         & $z_i$ & $a_i$ & $b_i$ & $c_i$ \\
         \hline\hline
         \multirow{10}{*}{$D^1_1$}& 1 & $0.6607 \cdot 10^{-7}$ & $-0.1767 \cdot 10^{-6}$ & $-0.1271 \cdot 10^{-4}$  \\
         \cline{2-5}
         & 2 & $-0.1750 \cdot 10^{-4}$ & $-0.5895 \cdot 10^{-5}$ &  $-0.8847 \cdot 10^{-6}$ \\
         \cline{2-5}
         & 3 & $-0.6507 \cdot 10^{-5}$& $-0.5144 \cdot 10^{-5}$ & $-0.1517 \cdot 10^{-5}$  \\
         \cline{2-5}
         & 4 & $-0.3851 \cdot 10^{-5}$& $0.1657 \cdot 10^{-4}$ & $-0.7559 \cdot 10^{-5}$ \\
         \cline{2-5}
         & 5 & $0.4987 \cdot 10^{-6}$& $0.2201 \cdot 10^{-5}$ & $-0.3089 \cdot 10^{-5}$  \\
         \cline{2-5}
         & 6 & $0.1953 \cdot 10^{-6}$ & $-0.3159 \cdot 10^{-4}$ & $-0.1827 \cdot 10^{-4}$  \\
         \cline{2-5}
         & 7 & $-0.2797 \cdot 10^{-4}$& $0.3893 \cdot 10^{-5}$ & $0.1194 \cdot 10^{-5}$  \\
         \cline{2-5}
         & 8 & $0.1604 \cdot 10^{-4}$& $-0.5762 \cdot 10^{-4}$ & $-0.1570 \cdot 10^{-4}$  \\
         \cline{2-5}
         & 9 & $-0.1183 \cdot 10^{-4}$& $-0.9919 \cdot 10^{-6}$ & $0.9930 \cdot 10^{-5}$ \\
         \cline{2-5}
         & 10 & $-0.7835 \cdot 10^{-5}$& $0.2179 \cdot 10^{-4}$ & $-0.7949 \cdot 10^{-5}$ \\
         \hline
    \end{tabular}\begin{tabular}{|c|c||c|c|c|}
        \hline
         & $z_i$ & $a_i$ & $b_i$ & $c_i$ \\
         \hline\hline
         \multirow{10}{*}{$D^2_0$}& 1 & $0.1980 \cdot 10^{-3}$ & $0.1510 \cdot 10^{-3}$ & $-0.2527 \cdot 10^{-4}$  \\
         \cline{2-5}
         & 2 & $-0.2572 \cdot 10^{-2}$& $-0.5907 \cdot 10^{-4}$ & $0.1149 \cdot 10^{-4}$  \\
         \cline{2-5}
         & 3 & $-0.2024 \cdot 10^{-2}$& $-0.2137 \cdot 10^{-4}$ & $0.1067 \cdot 10^{-4}$  \\
         \cline{2-5}
         & 4 & $0.1600 \cdot 10^{-2}$& $0.5689 \cdot 10^{-3}$ & $-0.2189 \cdot 10^{-3}$  \\
         \cline{2-5}
         & 5 & $0.1790 \cdot 10^{-3}$ & $0.1280 \cdot 10^{-3}$ & $-0.7452 \cdot 10^{-4}$  \\
         \cline{2-5}
         & 6 & $0.1228 \cdot 10^{-2}$& $-0.3551 \cdot 10^{-4}$ & $0.9342 \cdot 10^{-4}$  \\
         \cline{2-5}
         & 7 & $0.9168 \cdot 10^{-3}$& $-0.7961 \cdot 10^{-3}$ & $-0.1405 \cdot 10^{-3}$  \\
         \cline{2-5}
         & 8 & $0.4960 \cdot 10^{-2}$ & $-0.1981 \cdot 10^{-3}$ & $0.7174 \cdot 10^{-4}$  \\
         \cline{2-5}
         & 9 & $0.5225 \cdot 10^{-2}$& $0.2966 \cdot 10^{-3}$ & $-0.8173 \cdot 10^{-4}$  \\
         \cline{2-5}
         & 10 & $0.1550 \cdot 10^{-2}$& $-0.1694 \cdot 10^{-3}$ & $0.6938 \cdot 10^{-4}$  \\
         \hline
    \end{tabular}\\
    \begin{tabular}{|c|c||c|c|c|}
        \hline
         & $z_i$ & $a_i$ & $b_i$ & $c_i$ \\
         \hline\hline
         \multirow{10}{*}{$\theta_2$}& 1 & $-0.3160$ & $0.7038 \cdot 10^{-1}$ & $-0.2480 \cdot 10^{-1}$  \\
         \cline{2-5}
         & 2 & $-0.2355$& $0.2380 \cdot 10^{-1}$ & $0.6701 \cdot 10^{-2}$ \\
         \cline{2-5}
         & 3 & $-0.2021$& $0.1687 \cdot 10^{-1}$ & $0.5869 \cdot 10^{-2}$  \\
         \cline{2-5}
         & 4 & $0.4885 \cdot 10^{-1}$& $0.6057 \cdot 10^{-1}$ & $-0.2094 \cdot 10^{-1}$  \\
         \cline{2-5}
         & 5 & $-0.1106 \cdot 10^{-1}$& $0.2317 \cdot 10^{-1}$ & $-0.1128 \cdot 10^{-1}$ \\
         \cline{2-5}
         & 6 & $0.8406 \cdot 10^{-2}$ & $0.7702 \cdot 10^{-1}$ & $-0.2254 \cdot 10^{-1}$  \\
         \cline{2-5}
         & 7 & $0.3569 \cdot 10^{-2}$& $0.1531$ & $.1103$  \\
         \cline{2-5}
         & 8 & $0.3021 \cdot 10^{-1}$ & $0.1027 \cdot 10^{-2}$ & $-0.4945 \cdot 10^{-2}$  \\
         \cline{2-5}
         & 9 & $0.2762 \cdot 10^{-1}$& $0.2859 \cdot 10^{-2}$ & $-0.1297 \cdot 10^{-1}$  \\
         \cline{2-5}
         & 10 & $0.7229 \cdot 10^{-2}$& $0.1513 \cdot 10^{-1}$ & $0.1340 \cdot 10^{-1}$  \\
         \hline
    \end{tabular}

    \caption{Solutions for the Schenk parametrization by DFGS \cite{DFGS} }
    \label{table:app_schenk_descotes}
\end{table}

\section{$\overline{b}_{i}$ representation} \label{App_b_representation}
\begin{align}
x_{2} = &\frac{M_{\pi}^{2}}{F_{\pi}^{2}}\\
F^{(1)}(s) = &\frac{1}{2}J(s)\left[s^{2}-1\right]\\
G^{(1)}(s,t) = &\frac{1}{6}J(t)\left[14-4s-10t+st+2t^{2}\right]\\
\nonumber F^{(2)}(s) = &J(s)[\frac{1}{16\pi^{2}}\left(\frac{503}{108}s^{3}-\frac{929}{54}s^{2}+\frac{887}{27}s-\frac{140}{9}\right)+\\
\nonumber &+b_{1}\left(4s-3\right)+b_{2}(s^{2}+4s-4)+\\
&+\frac{b_{3}}{3}(8s^{3}-21s^{2}+48s-32)+\frac{b_{4}}{3}(16s^{3}-71s^{2}+112s-48)]+\\
\nonumber &+\frac{1}{18}K_{1}(s)\left[20s^{3}-119s^{2}+210s-135-\frac{9}{16}\pi^{2}(s-4)\right]+\\
\nonumber &+\frac{1}{32}K_{2}(s)[s\pi^{2}-24]+\frac{1}{9}K_{3}(s)[s^{2}-17s+9]\\
\nonumber G^{(2)}(s,t) = &J(t)\lbrace\frac{1}{16\pi^{2}}\left[\frac{412}{27}-\frac{s}{54}(t^{2}+5t+159)-t\left(\frac{267}{216}t^{2}-\frac{727}{108}t+\frac{1571}{108}\right)\right]+\\
\nonumber &+b_{1}(2-t)+\frac{b_{3}}{3}(t-4)(2t+s-5)-\frac{b_{3}}{6}(t-4)^{2}(3t+2s-8)+\\
\nonumber &+\frac{b_{4}}{6}\left(2s(3t-4)(t-4)-32t+40t^{2}-11t^{3}\right)\rbrace+\\
&\frac{1}{36}K_{1}(t)\left[174+8s-10t^{3}+72t^{2}-185t-\frac{\pi^{2}}{16}(t-4)(3s-8)\right]+\\
\nonumber &\frac{1}{9}K_{2}(t)\left[1+4s+\frac{\pi^{2}}{64}t(3s-8)\right]+\\
\nonumber &\frac{1}{9}K_{3}(t)\left[1+3st-s+3t^{2}-9t\right]+\frac{5}{3}K_{4}(t)\left[4-2s-t\right]
\end{align}
and the \emph{J}, $K_{1}$, $K_{2}$, $K_{3}$ and $K_{4}$ functions can be calculated as follows
\begin{align}
h(s)=&\frac{1}{N\sqrt{z}}\log\frac{\sqrt{z}-1}{\sqrt{z}+1},\hspace{1 cm} z=1-\frac{4}{s}, \hspace{1 cm} N=16\pi^{2}\\
\begin{pmatrix}
J\\
K_{1}\\
K_{2}\\
K_{3}
\end{pmatrix} =&\begin{pmatrix}
0&0&z&-4N\\
0&z&0&0\\
0&z^{2}&0&8\\
Nzs^{-1}&0&\pi^{2}(Ns)^{-1}&\pi^{2}
\end{pmatrix}\begin{pmatrix}
h^{3}\\
h^{2}\\
h\\
\frac{-1}{2N^{2}}
\end{pmatrix}\\
K_{4}=&\frac{1}{sz}\left(\frac{1}{2}K_{1}+\frac{1}{3}K_{3}+\frac{J}{N}+\frac{(\pi^{2}-6)s}{12N^{2}}\right).
\end{align}

\section{Relations between amplitude representations} \label{App_repre_rel}
As shown in CGL \cite{CGL}, the chiral and phenomenological representations' nonpolynomial parts match up to $O(p^{6})$ if the polynomial parts relate such that
\begin{align}
\label{ci_pi}\nonumber c_{1}=&16\pi a_{0}^{2}+p_{1}+O(p^{8})\\ \nonumber
c_{2}=&\frac{4\pi}{3M_{\pi}^{2}}(2a_{0}^{0}-5a_{0}^{2})+p_{2}+O(p^{6})\\
\nonumber c_{3}=&p_{3}+O(p^{4})\\
c_{4}=&p_{4}+O(p^{4})\\
\nonumber c_{5}=&p_{5}+O(p^{2})\\
\nonumber c_{6}=&p_{6}+O(p^{2}). 
\end{align}

Next are the transformations between $c_{i}$ and $\overline{b}_{i}$'s
\begin{align}
\label{ci_bi}\nonumber c_{1}=&-\frac{M_{\pi}^{2}}{F_{\pi}^{2}}\left[1+\xi\left(-\overline{b}_{1}-\frac{68}{315}\right)+\xi^{2}\left(-\frac{8\overline{b}_{1}}{105}-\frac{32\overline{b}_{2}}{63}-\frac{464\overline{b}_{3}}{315}-\frac{3824\overline{b}_{4}}{315}+\frac{601\pi^{2}}{945}-\frac{17947}{2835}\right)\right]\\ 
\nonumber c_{2}=&\frac{1}{F_{\pi}^{2}}\left[1+\xi\left(\overline{b}_{2}-\frac{323}{1260}\right)+\xi^{2}\left(-\frac{11\overline{b}_{1}}{70}-\frac{211\overline{b}_{2}}{315}-\frac{628\overline{b}_{3}}{315}-\frac{5164\overline{b}_{4}}{315}-\frac{3977}{630}+\frac{5237\pi^{2}}{7560}\right)\right]\\
\nonumber c_{3}=&\frac{1}{N^{2}F_{\pi}^{4}}\left(\overline{b}_{3}+\frac{1}{42}+\xi\left(\frac{18\overline{b}_{1}}{35}+\frac{59\overline{b}_{2}}{105}+\frac{731\overline{b}_{3}}{315}+\frac{3601\overline{b}_{4}}{315}-\frac{5387\pi^{2}}{15120}-\frac{19121}{7560}\right)\right]\\
c_{4}=&\frac{1}{N^{2}F_{\pi}^{4}}\left(\overline{b}_{4}-\frac{31}{2520}+\xi\left(-\frac{43\overline{b}_{1}}{420}-\frac{8\overline{b}_{2}}{63}+\frac{23\overline{b}_{3}}{63}+\frac{997\overline{b}_{4}}{315}+\frac{467\pi^{2}}{7560}-\frac{63829}{45360}\right)\right]\\ \nonumber
c_{5}=&\frac{1}{N^{2}F_{\pi}^{6}}\left(\frac{137}{1680\xi}+\frac{\overline{b}_{1}}{16}+\frac{379\overline{b}_{2}}{1680}-\frac{25\overline{b}_{3}}{28}-\frac{731\overline{b}_{4}}{180}+\overline{b}_{5}+\frac{269\pi^{2}}{15120}+\frac{61673}{18144}\right)\\
\nonumber c_{6}=&\frac{1}{N^{2}F_{\pi}^{6}}\left(-\frac{31}{1680\xi}+\frac{\overline{b}_{1}}{112}-\frac{47\overline{b}_{2}}{1680}-\frac{65\overline{b}_{3}}{252}-\frac{547\overline{b}_{4}}{420}+\overline{b}_{6}+\frac{\pi^{2}}{15120}+\frac{44287}{90720}\right),
\end{align}
where $\xi=\left(\frac{M_{\pi}}{4\pi F_{\pi}}\right)^{2}$ and $N=16\pi^{2}$.
\end{document}